\documentclass[12pt]{article}

\usepackage{amsmath}
\usepackage{amssymb}
\usepackage{graphicx}
\usepackage{cite}
\usepackage{hyperref}

\begin{document}
\title{Gravitational waves from bubble walls}
\author{
Ariel M\'{e}gevand\thanks{Member of CONICET, Argentina. E-mail address:
megevand@mdp.edu.ar}~ 
and Federico Agust\'{\i}n Membiela\thanks{Member of CONICET, Argentina. E-mail
address: membiela@mdp.edu.ar} \\[0.5cm]
\normalsize \it IFIMAR (CONICET-UNMdP)\\
\normalsize \it Departamento de F\'{\i}sica, Facultad de Ciencias Exactas
y Naturales, \\
\normalsize \it UNMdP, De\'{a}n Funes 3350, (7600) Mar del Plata, Argentina }
\date{}
\maketitle
\begin{abstract}
We  present a general method for computing the gravitational radiation arising from the motion
of bubble walls or thin fluid shells in cosmological phase transitions.
We discuss the application of  this method to different wall kinematics.
In particular, we derive general expressions for the 
bubble collision mechanism in the envelope approximation and the so-called bulk flow model, 
and we also consider deformations from the spherical bubble shape.
We calculate the gravitational wave spectrum for a specific model of deformations on a definite size scale, 
which gives a peak away from that of the bubble collision mechanism.
\end{abstract}

\section{Introduction}

A phase transition of the Universe may give rise to  gravitational waves (GWs) \cite{tw90}. In particular,
a first-order phase transition occurring at the electroweak scale gives naturally a GW spectrum which may
be observable by the space-based interferometer LISA \cite{LISA}. 
A first-order phase transition occurs via the nucleation
and expansion of bubbles of a stable phase into a metastable one.
In general, the phase transition is modeled with a scalar
order-parameter field $\phi(\mathbf{x},t)$ which couples to a relativistic fluid 
representing the plasma. 
A bubble corresponds to a configuration
in which the scalar field takes the stable-phase value in some region,
while outside this region it takes the value corresponding 
to the metastable phase (usually, $\phi=0$). 
The interfaces or bubble walls propagate in the hot plasma. 
In many cases, they reach a terminal 
velocity due to the friction with the plasma 
(see, e.g., \cite{lmt92,t92,dlhll92,k92,a93,mp95l,mp95}).
Alternatively, the walls may exhibit runaway behavior \cite{bm09,bm17,av21} and propagate 
with velocity $ v=1 $ like in a vacuum phase transition.
On the other hand, it has been shown that walls which propagate as subsonic deflagrations
are unstable below a certain critical velocity
\cite{l92,hkllm93,mm14,mms15}.
This instability causes the exponential growth of wall deformations as well as turbulent fluid motions near the walls.
 
Thus, the bubble walls may generate
GWs in several ways. In the bubble collision mechanism \cite{ktw92,ktw92b},
the gravitational radiation is produced directly by the wall motion. 
Other mechanisms, such as turbulence \cite{kkt94,dgn02,kmk02,gkk07,cds09,kks10,kk15,nss18,pmbkk20}
and sound waves \cite{hhrw14,gm14,hhrw15,hhrw17,h18,hh19,gsvw21}, originate
in the bulk fluid motions caused by the walls. 
Neither of these mechanisms take into account the possibility of wall deformations or 
the fluid motions caused by hydrodynamic instabilities.

The original calculation method for the bubble-collision mechanism
is called the envelope approximation \cite{kt93}. This approach consists
in simulating the nucleation and expansion of spherical bubbles and considering only their walls 
as sources of GWs. The walls are assumed to be infinitely-thin and to disappear where bubbles overlap.
The expansion
of these spheres is followed until the end of the phase transition.
The envelope approximation has also been used to compute the gravitational
radiation from bulk fluid motions, assuming that the fluid is concentrated
in thin shells next to the walls \cite{kkt94,hk08}. 
In Ref.~\cite{jt17} it was pointed out that most of the calculations
with the envelope approximation can be done analytically. The technique
is based on a previous analytic approach \cite{cds08}, where the
ensemble average of the source is taken. 
A similar approach
was used more recently to consider thin fluid shells which persist after the walls
collide \cite{jt19}. This modification of the envelope approximation
was called the bulk flow model in \cite{k18}, where it was considered
with the approach of simulating the formation and expansion of the
thin fluid shells.

In the present paper we introduce a general method for calculating the GW spectrum
from thin walls or fluid shells, which allows to consider different wall dynamics.
Our approach is a hybrid between the one used in Refs.~\cite{kt93,hk08}
and that of Refs.~\cite{jt17,cds08}. To illustrate the method,
we derive expressions for the envelope approximation and
the bulk flow model, which we compare with previous approaches.
We also derive general expressions for arbitrary
wall deformations and consider a particular example,
in which we model deformations on a given length scale.
This introduces a peak in the GW spectrum at a higher frequency than the usual generation mechanisms.

The plan is as follows.
In the next section we introduce our general approach to the calculation of gravitational
waves.
In Sec.~\ref{bbcoll} we discuss the envelope approximation. 
In Sec.~\ref{bfm} we
consider the bulk flow model.
In Sec.~\ref{wdef} we address the case of wall deformations,
and in Sec.~\ref{toymodel} we discuss a specific model.
We summarize our conclusions in Sec.~\ref{conclu}. 
A detailed comparison of our approach with previous ones can be found in appendix \ref{approaches}, while  
appendices \ref{angint}-\ref{apfaseestac} contain details on the calculations.

\section{Bubble kinematics and gravitational waves}
\label{gwwalls}

It is reasonable to assume that bubbles are initially spherical. However, 
since a spherical source does not radiate, a bubble must lose this symmetry in order to produce GWs.
In the envelope approximation or in the bulk flow model, this happens as the bubbles collide and
the bubble walls or fluid shells 
carry different surface energy density in their collided and uncollided parts. 
In contrast, a non-spherical bubble does not need to be collided to generate GWs.
Besides the already mentioned hydrodynamic instabilities, deformations can arise due to bubble collisions
or other interactions.
In particular, the inhomogeneous reheating due to shock fronts preceding deflagration fronts may cause velocity variations
along a given bubble wall.

On the other hand, for the development of the phase transition it
is simpler to model the bubbles as overlapping spheres, and we shall use this approximation
(we shall take into account deformations only as a source of GWs).
Assuming a homogeneous rate $\Gamma(t)$
per unit time per unit volume and a homogeneous wall velocity $v(t)$,
a relatively simple statistical treatment of the phase transition
is possible \cite{gt80,gw81,tww92}. We shall ignore for simplicity
the scale factor (which is valid if the transition is short enough),
and we shall also make the usual approximation of neglecting the initial
bubble size. Then, the radius of a bubble which nucleated at time $t_{N}$
and expanded until time $t$ is given by 
\begin{equation}
R(t_{N},t)=\int_{t_{N}}^{t}v(t'')dt''.\label{Rgral}
\end{equation}
Taking into account bubble
overlapping, the average fraction of volume remaining in the high-temperature
phase at time $t$ is given by $f_{+}(t)=e^{-I(t)}$, with 
\begin{equation}
I(t)=\int_{-\infty}^{t}dt''\Gamma(t'')\frac{4\pi}{3}R(t'',t)^{3}.\label{It}
\end{equation}
The lower limit of integration can be replaced with the time at which the critical temperature is reached,
since the nucleation rate vanishes before that.

The relevant quantity in the computation of gravitational waves is
the stress-energy tensor $T_{ij}(t,\mathbf{x})$ of the source. For
the mechanism of bubble collisions in the envelope approximation,
there have been essentially two different approaches. In numerical
simulations such as those of Refs.~\cite{kt93,hk08}, the radiated
energy is considered in the wave zone approximation. The stress-energy
tensor is decomposed as a sum over bubbles, and the Fourier transform
of each contribution is considered. A certain number of bubbles is
nucleated, and then the result is summed and averaged over realizations
of the phase transition. On the other hand, in the semi-analytic approach
introduced in Refs.~\cite{cds08,jt17}, the evolution of the metric
perturbations in Fourier space is calculated and linked to the
source, and then the ensemble average is taken. This calculation involves
the Fourier transform of the average $\left\langle T_{ij}(t,\mathbf{x})T_{kl}(t',\mathbf{x}')\right\rangle $,
and this quantity is related to the probability for the presence of
bubble walls at the points $(t,\mathbf{x})$, $(t',\mathbf{x}')$.
The two approaches are of course equivalent. We compare them in some
detail in appendix \ref{approaches}. Here we shall use a hybrid approach.
We shall decompose the stress-energy tensor into individual bubbles
like in Ref.~\cite{kt93}, but we shall consider analytically the ensemble
average. This method will allow us to obtain general results in fewer
steps, and is also suitable for considering more general wall dynamics
than that of bubble expansion.

We begin by writing the gravitational wave power spectrum for a
large volume $V$ in the form 
\begin{equation}
\frac{d\rho_{GW}}{d\ln\omega}=\frac{4G\omega^{3}}{\pi}\int_{-\infty}^{\infty}dt\int_{t}^{\infty}dt'\cos[\omega(t-t')]\,\Pi(t,t',\omega),\label{rhogw}
\end{equation}
where
\begin{equation}
\Pi(t,t',\omega)\equiv\frac{1}{V}\Lambda_{ij,kl}(\hat{n})\left\langle \tilde{T}_{ij}(t,\omega\hat{n})\tilde{T}_{kl}(t',\omega\hat{n})^{*}\right\rangle .\label{defPi}
\end{equation}
Here, the symbol $\langle\ \rangle$ denotes ensemble average, $\Lambda_{ij,kl}(\hat{n})$
is the transverse-traceless projection tensor
\begin{equation}
\Lambda_{ij,kl}=\delta_{ik}\delta_{jl}-\frac{1}{2}\delta_{ij}\delta_{kl}-\hat{n}_{j}\hat{n}_{l}\delta_{ik}-\hat{n}_{i}\hat{n}_{k}\delta_{jl}+\frac{1}{2}\hat{n}_{k}\hat{n}_{l}\delta_{ij}+\frac{1}{2}\hat{n}_{i}\hat{n}_{j}\delta_{kl}+\frac{1}{2}\hat{n}_{i}\hat{n}_{j}\hat{n}_{k}\hat{n}_{l},
\end{equation}
$\hat{n}$ is the direction of observation, and $\tilde{T}_{ij}$
are the spatial components of the Fourier transform of the stress-energy
tensor,
\begin{equation}
\tilde{T}_{ij}(t,\omega\hat{n})=\int_{V}d^{3}xe^{-i\omega\hat{n}\cdot\mathbf{x}}T_{ij}(t,\mathbf{x}).\label{defTijtil}
\end{equation}
In appendix \ref{approaches} we derive these formulas and compare
with the aforementioned approaches. We also show that homogeneity
and isotropy imply that the quantity $\Pi(t,t',\omega)$ is real,
does not depend on $\hat{n}$, and satisfies $\Pi(t,t',\omega)=\Pi(t',t,\omega)$.

From now on we shall denote the total energy-momentum tensor of the
system as $T_{ij}^{\mathrm{tot}}$, while $T_{ij}$ will denote the
contribution from a single bubble. Each bubble gives rise to different
GW sources (the wall motion, bulk fluid motions, etc.). We shall consider
cases in which the GW sources which originate in different bubbles
do not overlap, so that the total energy-momentum tensor is of the
form
\begin{equation}
T_{ij}^{\mathrm{tot}}(t,\mathbf{x})=\sum_{n}T_{ij}^{(n)}(r_{n},\hat{r}_{n},t),\label{decomp}
\end{equation}
where $r_{n}=|\mathbf{x}-\mathbf{x}_{n}|$, $\hat{r}_{n}=(\mathbf{x}-\mathbf{x}_{n})/r_{n}$,
and $\mathbf{x}_{n}$ is the position of the center of the $n$th
bubble. In this description, we define the bubble as a sphere centered
at $\mathbf{x}_{n}$ and with a radius $R(t_{N}^{(n)},t)$ given by
Eq.~(\ref{Rgral}), where $t_{N}^{(n)}$ is the nucleation time.
However, the support of the associated contribution $T_{ij}^{(n)}$
might be larger or smaller than this sphere, depending on the source
we consider.

Inserting Eq.~(\ref{decomp}) in Eq.~(\ref{defTijtil}), we obtain
\begin{equation}
\tilde{T}_{ij}^{\mathrm{tot}}(t,\omega\hat{n})=
\sum_{n}e^{-i\omega\hat{n}\cdot\mathbf{x}_{n}}\tilde{T}_{ij}^{(n)}(t,\omega\hat{n}),
\end{equation}
with 
\begin{equation}
\tilde{T}_{ij}^{(n)}(t,\omega\hat{n})=\int r^{2}dr\int d\hat{r}e^{-i\omega\hat{n}\cdot\mathbf{r}}T_{ij}^{(n)}(r,\hat{r},t),\label{Ttilden}
\end{equation}
where the origin of the radial variable $r$ is the center of the
$n$th bubble, and we use the notation $d\hat{r}\equiv\sin\theta d\theta d\phi$.
We have to compute the quantity (\ref{defPi}),
\begin{equation}
\Pi(t,t',\omega)=\frac{1}{V}\left\langle \sum_{n}\sum_{m}e^{-i\omega\hat{n}\cdot(\mathbf{x}_{n}-\mathbf{x}_{m})}\Lambda_{ij,kl}(\hat{n})\tilde{T}_{ij}^{(n)}(t,\omega\hat{n})\tilde{T}_{kl}^{(m)}(t',\omega\hat{n})^{*}\right\rangle .\label{Pisuma}
\end{equation}
It is convenient to calculate separately the contributions of $m=n$
and those of $m\neq n$, since the correlations within a given bubble
are different from those between different bubbles. Moreover, both
contributions scale with the volume $V$ \cite{mm20}, in spite of
the fact that the number of terms in the double sum scales with $V^{2}$.
This naturally decomposes Eq.~(\ref{Pisuma}) into single-bubble
and two-bubble contributions, and we shall use indices $s,d$ for
them, like in the treatment of Ref.~\cite{jt17}.

Let us first consider the single-bubble contribution, 
\begin{equation}
\Pi^{(s)}(t,t',\omega)=\Lambda_{ij,kl}(\hat{n})\frac{1}{V}\left\langle \sum_{n}\tilde{T}_{ij}^{(n)}(t,\omega\hat{n})\tilde{T}_{kl}^{(n)}(t',\omega\hat{n})^{*}\right\rangle .\label{Pisumas}
\end{equation}
For each term,  $t$ and $t'$ are two instants in the history of the same bubble
(labeled by $n$), which was nucleated at some time $t_{N}^{(n)}$ previous to
both $t$ and $t'$. We can think of the ensemble as a large number
of realizations of the phase transition. For each realization, we
may arrange the terms of the sum in (\ref{Pisumas}) into groups of
bubbles nucleated in the same neighborhood around the same time. Thus,
the average in Eq.~(\ref{Pisumas}) separates into a sum of averages,
each of which involves only bubbles nucleated in a certain region
and time. The group of bubbles nucleated around a given position $\mathbf{x}_{N}$
and in a time interval $[t_{N},t_{N}+dt_{N}]$ will have, on average,
a number of members $dN_{b}$ given by 
\begin{equation}
dN_{b}=\Gamma(t_{N})f_{+}(t_{N})d^{3}x_{N}dt_{N}.
\end{equation}
If we pick one bubble of this group from each realization of the phase
transition, we may use such a collection to formally define the average
of the quantity $\tilde{T}_{ij}\tilde{T}_{kl}^{*}$ for a single bubble.
By homogeneity, the resulting average does not depend on the bubble
position $\mathbf{x}_{N}$, and we denote it as\footnote{Notice that $\tilde{T}_{ij}$ is given by Eq.~(\ref{Ttilden}), where
the bubble index must be dropped when taking the average.}
\begin{equation}
\left\langle \tilde{T}_{ij}(t,\omega\hat{n})\tilde{T}_{kl}(t',\omega\hat{n})^{*}\right\rangle ^{(s)}(t_{N}).\label{singlecontrib}
\end{equation}
Multiplying by the average number $dN_{b}$ and integrating over the
nucleation position, the volume factor in Eq.~(\ref{Pisumas}) cancels
and we obtain
\begin{equation}
\Pi^{(s)}(t,t',\omega)=\int_{-\infty}^{t}dt_{N}\Gamma(t_{N})f_{+}(t_{N})\Lambda_{ij,kl}(\hat{n})\left\langle \tilde{T}_{ij}(t,\omega\hat{n})\tilde{T}_{kl}(t',\omega\hat{n})^{*}\right\rangle ^{(s)}.\label{Pis}
\end{equation}
The nucleation time must be integrated from $-\infty$ (or from $t_{c}$)
to the smallest of $t$ and $t'$, which, according to Eq.~(\ref{rhogw})\footnote{If we use instead the symmetric expression (\ref{rhogwTF}) of appendix
\ref{approaches} for $d\rho_{GW}/d\ln\omega,$ the limit of integration
in (\ref{Pis}) must be replaced by $t_{m}=\min\{t,t'\}$.}, is $t$.

Now let us consider the two-bubble contribution, which is given by
Eq.~(\ref{Pisuma}) with $n\neq m$. We may divide each sum into
groups of bubbles nucleated in small neighborhoods around positions
$\mathbf{x}_{N}$ and $\mathbf{x}_{N}'$ and in small time intervals
around $t_{N}$ and $t_{N}'$. However, by homogeneity, the contribution
of a pair of neighborhoods to the average will depend only on the
relative position $\mathbf{l}=\mathbf{x}_{N}'-\mathbf{x}_{N}$. If
we pick, from each realization of the phase transition, one such pair
of bubbles and average the quantity $\tilde{T}_{ij}\tilde{T}_{kl}^{*}$,
we obtain the two-bubble ensemble average
\begin{equation}
\left\langle \tilde{T}_{ij}(t,\omega\hat{n})\tilde{T}_{kl}(t',\omega\hat{n})^{*}\right\rangle ^{(d)}(t_{N},t_{N}',\mathbf{l}).
\end{equation}
To calculate the double sum of Eq.~(\ref{Pisuma}), we must integrate
over $\mathbf{x}_{N}$ and $\mathbf{x}_{N}'$, which gives a factor
of $V$ and an integral over $\mathbf{l}$. The volume factor cancels,
and we obtain 
\begin{align}
\Pi^{(d)}(t,t',\omega)= & \int_{-\infty}^{t}dt_{N}\Gamma(t_{N})f_{+}(t_{N})\int_{t_{c}}^{t'}dt_{N}^{\prime}\Gamma(t_{N}^{\prime})f_{+}(t_{N}^{\prime})\nonumber \\
 & \times\int d^{3}l\,e^{i\omega\hat{n}\cdot\mathbf{l}}\Lambda_{ij,kl}(\hat{n})\left\langle \tilde{T}_{ij}(t,\omega\hat{n})\tilde{T}_{kl}(t',\omega\hat{n})^{*}\right\rangle ^{(d)}.\label{Pid}
\end{align}
Notice that, in this expression, the contribution of correlations
between bubbles of the same age is negligible, since it is of order
$(dN_{b})^{2}$. Nevertheless, single-bubble correlations give the finite
contribution (\ref{Pis}).

The expressions (\ref{Pis}) and (\ref{Pid}) will be valid whenever
the quantities $T_{ij}$ belonging to different bubbles have little
overlapping. For instance, in the case of hydrodynamic instabilities
which may be developed by deflagration walls, unstable modes grow
exponentially in the surrounding fluid (besides a corrugation of the
wall). These perturbations break the spherical symmetry, so GWs may
be produced well before bubbles begin to overlap. On the other hand,
if the energy-momentum tensor is concentrated in thin shells, the
overlapping will be negligible even when the bubbles overlap significantly.

\section{The envelope approximation}

\label{bbcoll}

A bubble is a configuration in which the scalar field $\phi$ has
a non-vanishing constant value inside a certain domain and falls to
zero outside this region. The non-diagonal components of $T_{ij}$
for the scalar field are given by $T_{ij}^{\phi}=\partial_{i}\phi\partial_{j}\phi$,
and are non-vanishing only where $\phi$ has a spatial variation,
which is, by definition, the bubble wall. For a spherical configuration
of the form $\phi=\phi(r)$, we have $T_{ij}^{\phi}=\hat{r}_{i}\hat{r}_{j}(\partial_{r}\phi)^{2}$.
More specifically, we shall write the field profile of the bubble
in the form 
\begin{equation}
\phi(\mathbf{x},t)=\phi_{0}(r-R),\label{perfesf}
\end{equation}
where $\phi_{0}$ is the solution of the 1-dimensional problem (a
$\tanh$ profile is often used as an approximation), and $R\equiv R(t_{N},t)$
is the position of the wall, which is given by Eq.~(\ref{Rgral}). We shall
use the approximation of infinitely thin walls, which we implement
by writing $\phi_{0}^{\prime}(x)^{2}=\sigma_{0}\delta(x)$,
where $\phi_{0}^{\prime}\equiv d\phi_{0}/dx$, and $\sigma_{0}=\int\phi_{0}^{\prime2}dx$
is the surface tension. This quantity can be computed from the field
profile, and we shall regard it as a free parameter. In the envelope
approximation, we assume that the wall just disappears
in the intersections with other bubbles. Thus, we have 
\begin{equation}
T_{ij}^\phi=\hat{r}_{i}\hat{r}_{j}\,\sigma_{0}\delta(r-R)\,1_{S}(\hat{r}),\label{Tijenv}
\end{equation}
where $1_{S}$ is the indicator function for the uncollided bubble
surface $S$ at time $t$,
\begin{equation}
1_{S}(\hat{r})=\begin{cases}
1 & \text{if \ensuremath{R\hat{r}} is uncollided,}\\
0 & \text{otherwise.}
\end{cases}\label{indicator}
\end{equation}
Therefore, the quantity $\tilde{T}_{ij}$ defined by Eq.~(\ref{Ttilden})
becomes
\begin{equation}
\tilde{T}_{ij}(t,\omega\hat{n})=\sigma_{0}R^{2}\int d\hat{r}e^{-i\omega R\hat{n}\cdot\hat{r}}\hat{r}_{i}\hat{r}_{j}1_{S}(\hat{r}).\label{Ttilijenv}
\end{equation}

We shall now replace the constant $\sigma_{0}$ by a variable surface
energy density $\sigma$ in order to take into account the fact that
energy may accumulate in the wall. To obtain this replacement formally,
we may replace Eq.~(\ref{perfesf}) by $\phi(\mathbf{x},t)=\phi_{0}(\gamma(r-R))$,
where $\gamma=1/\sqrt{1-v^{2}}$, so that the contraction of the wall
width is taken into account (see, e.g., \cite{m13,lm16}). This 
gives $ (\partial_r \phi)^2=\gamma^2\sigma_0\delta\left(\gamma(r-R)\right) $,
which 
results in a single factor
$\gamma$ in Eq.~(\ref{Ttilijenv}), corresponding to a surface energy
density $\sigma=\gamma\sigma_{0}$. 
This factor grows with time if the wall is accelerated\footnote{It can be seen \cite{kt93} that, 
	in the relativistic limit, the kinetic
	and gradient energy terms inside the wall, $\frac{1}{2}(\partial_{r}\phi)^{2}$,
	$\frac{1}{2}(\partial_{t}\phi)^{2}$, become equal in a short time,
	and both terms grow with $t^{3}$ while the potential energy becomes
	negligible. Therefore $\sigma=\int(\partial_{r}\phi)^{2}dr=\gamma\sigma_{0}$
	gives indeed the energy density which is accumulated inside the wall.}.
In a vacuum transition, all of the released energy goes to the wall,
so the surface energy $4\pi R^{2}\sigma$ is given by the released
energy $(4\pi/3)R^{3}\rho_{\mathrm{vac}}$. This gives $\sigma=(\rho_{\mathrm{vac}}/3)R$.

In contrast, if the phase transition occurs in a hot plasma, the wall
may reach a constant velocity. In this case, most of the released
energy goes to the fluid, and we have $\sigma=\sigma_{0}$, which
will give a much lower GW signal\footnote{Dimensionally, we have $\sigma_{0}\sim T^{3}$, $\rho_{\mathrm{vac}}\sim T^{4}$,
and $R\sim M_{P}/T^{2}$, where $M_{P}$ is the Planck mass. Hence there is
an enhancement of order $M_{P}/T$ for $\sigma=(\rho_{\mathrm{vac}}/3)R$
with respect to $\sigma=\sigma_{0}$.}. Nevertheless, a fluid profile forms next to the bubble wall, and
the envelope approximation has been used for this case, assuming that
the energy is concentrated in a thin shell.
For the fluid, the relevant part of the energy-momentum tensor is
$T_{ij}^{\mathrm{fl}}=w\gamma^{2}v_{\mathrm{fl}i}v_{\mathrm{fl}j}$,
where $w$ is the enthalpy density, $v_{\mathrm{fl}}$ is the fluid
velocity, and $\gamma_{\mathrm{fl}}=1/\sqrt{1-v_{\mathrm{fl}}^{2}}$.
For a spherical configuration, the velocity is of the form $\mathbf{v}_{\mathrm{fl}}=v_{\mathrm{fl}}(r)\hat{r}$
and the enthalpy density is of the form $w=w(r)$. Therefore, we have
$T_{ij}^{\mathrm{fl}}=\hat{r}_{i}\hat{r}_{j}w\gamma_{\mathrm{fl}}^{2}v_{\mathrm{fl}}^{2}$,
which is of the same form of $T_{ij}^{\phi}$, namely, $f(r)\hat{r}_{i}\hat{r}_{j}$,
although the function $f(r)$ has a wider support for the fluid. 
Assuming
anyway a thin fluid shell, the approximation (\ref{Ttilijenv}) can
be used, with $ \sigma_0 $ replaced by
\begin{equation}
\sigma=(\kappa\rho_{\mathrm{vac}}/3)R,\label{sigmarhovac}
\end{equation}
where $\kappa$ is an efficiency factor accounting for the fraction
of energy which goes to bulk fluid motions \cite{kkt94}. It can be
calculated from the fluid velocity profile (see, e.g., \cite{m08,ekns10,lm11,lm15,lm16,elnv19,elv20,gkv20,gkv21}), and we shall consider
it as a free parameter here. In an intermediate case in which the wall
is accelerated but a part of the energy goes to the fluid, there are
different efficiency factors for the wall and for the fluid \cite{lm16b,gwlisa}.
In the envelope approximation, $\kappa$ represents the sum of these
contributions\footnote{It is worth noticing that the usual assumption that $\kappa$ and
$\rho_{\mathrm{vac}}$ are constant parameters is an approximation (which we adopt here).
The released energy $\rho_{\mathrm{vac}}$ depends on the temperature
(which varies during the phase transition), while the efficiency factor
$\kappa$ depends also on the wall velocity. To take into account
a varying transfer of energy to the wall, the released energy $\kappa\rho_{\mathrm{vac}}\frac{4\pi}{3}R^{3}$
should be replaced by $\int_{t_{N}}^{t}4\pi R^{2}\kappa\rho_{\mathrm{vac}}vdt''$.}.

\begin{figure}[tb]
\centering
\includegraphics[width=1\textwidth]{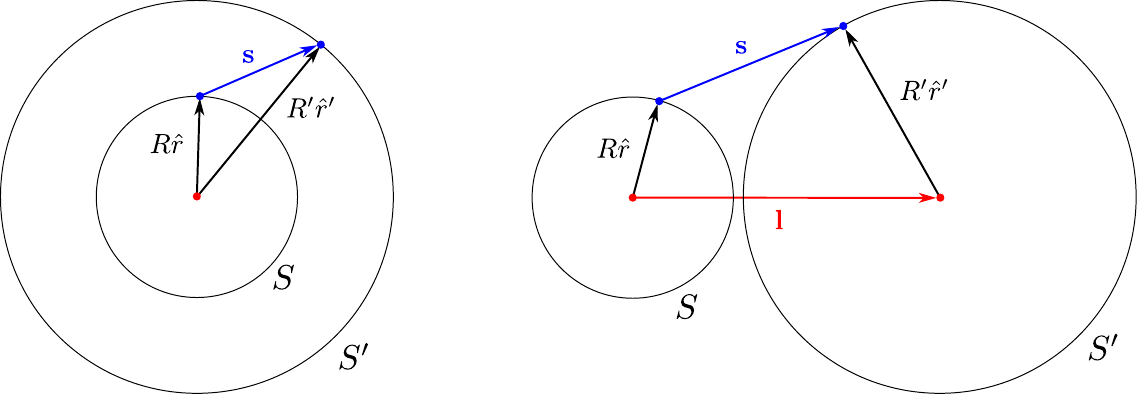}
\caption{Relations between the vectors $R\hat{r}$, $R'\hat{r}'$, $\mathbf{s}$,
and $\mathbf{l}$. Left: single-bubble case. Right: two-bubble case.}
\label{figdist}
\end{figure}

Inserting Eq.~(\ref{Ttilijenv}) in Eqs.~(\ref{Pis}) and (\ref{Pid}),
we obtain 
\begin{equation}
\Pi^{(s)}=\int_{-\infty}^{t}dt_{N}\Gamma(t_{N})f_{+}(t_{N})\sigma R^{2}\sigma'R^{\prime2}\int d\hat{r}\int d\hat{r}'\,e^{i\omega\hat{n}\cdot\mathbf{s}}\Lambda_{ij,kl}\hat{r}_{i}\hat{r}_{j}\hat{r}'_{k}\hat{r}'_{l}\left\langle 1_{S}(\hat{r})1_{S'}(\hat{r}')\right\rangle \label{Pisfinal-env}
\end{equation}
and
\begin{multline}
\Pi^{(d)}=\int_{-\infty}^{t}dt_{N}\Gamma(t_{N})f_{+}(t_{N})\sigma R^{2}\int_{-\infty}^{t'}dt_{N}^{\prime}\Gamma(t_{N}^{\prime})f_{+}(t_{N}^{\prime})\sigma'R^{\prime2}\\
\times\int d^{3}l\int d\hat{r}\int d\hat{r}'\,e^{i\omega\hat{n}\cdot\mathbf{s}}\Lambda_{ij,kl}\hat{r}_{i}\hat{r}_{j}\hat{r}'_{k}\hat{r}'_{l}\left\langle 1_{S}(\hat{r})1_{S'}(\hat{r}')\right\rangle .\label{Pidfinal-env}
\end{multline}
Here, $R'$ means $R(t_{N},t')$ for the single-bubble contribution
and $R(t_{N}^{\prime},t')$ for the two-bubble contribution, while
$S'$ is the uncollided surface of the corresponding bubble at time
$t'$. The vector $\mathbf{s}=\mathbf{l}+R'\hat{r}'-R\hat{r}$ (with
$\mathbf{l}=\mathbf{0}$ in the single-bubble case) connects the points
$p$ and $p'$ on the surfaces $S$ and $S'$ which are determined
by the directions $\hat{r}$ and $\hat{r}'$, respectively (see Fig.~\ref{figdist}).
The quantity $\Lambda_{ij,kl}\hat{r}_{i}\hat{r}_{j}\hat{r}'_{k}\hat{r}'_{l}$
can be written as 
\begin{align}
\Lambda(\hat{n})_{ij,kl}\hat{r}_{i}\hat{r}_{j}\hat{r}'_{k}\hat{r}'_{l}= & (\hat{r}\cdot\hat{r}')^{2}-\frac{1}{2}-2(\hat{r}\cdot\hat{r}')(\hat{n}\cdot\hat{r})(\hat{n}\cdot\hat{r}')\nonumber \\
 & +\frac{1}{2}(\hat{n}\cdot\hat{r}')^{2}+\frac{1}{2}(\hat{n}\cdot\hat{r})^{2}+\frac{1}{2}(\hat{n}\cdot\hat{r})^{2}(\hat{n}\cdot\hat{r}')^{2}.
 \label{Ltt0}
\end{align}
On the other hand, the product $1_{S}(\hat{r})1_{S'}(\hat{r}')$ takes
the value $1$ if the points $p$ and $p'$ on each spherical surface
are both uncollided and vanishes otherwise, so we have 
\begin{equation}
\left\langle 1_{S}(\hat{r})1_{S'}(\hat{r}')\right\rangle =P_{uu}(\hat{r},\hat{r}'),\label{SSPNC}
\end{equation}
where $P_{uu}$ is the probability that both points are uncollided
(at the corresponding times $t$ and $t'$).

The probability $P_{uu}$ was considered in Ref.~\cite{mm20} (see
\cite{jt17} for related calculations). The result is different if
the two points belong to the same bubble surface or to two different
bubbles. In the single-bubble case, we have 
\begin{equation}
P_{uu}^{(s)}=\exp\left[-I(t)-I(t')+I_{\cap}(t,t',s)+I(t_{N})\right],\label{Pus}
\end{equation}
where the function $I(t)$ is defined in Eq.~(\ref{It}), and the
quantity $I_{\cap}$ is given (for $t<t'$) by 
\begin{equation}
I_{\cap}(t,t',s)=\int_{-\infty}^{t}dt''\Gamma(t'')V_{\cap}(t'',t,t',s).\label{Itts}
\end{equation}
Here, $V_{\cap}$ is the volume of the intersection of two spheres
centered at the points $ p $ and $ p' $,
of radii $r=R(t'',t)$ and $r'=R(t'',t')$, respectively. Their centers are separated
by a distance $s=|\mathbf{s}|$, and the intersection volume 
is given by 
\begin{equation}
V_{\cap}=\frac{\pi}{12}(r+r'-s)^{2}\left[s+2(r+r')-\frac{3(r-r')^{2}}{s}\right]\Theta(r+r'-s),\label{VI}
\end{equation}
where $\Theta$ is the Heaviside step 
function\footnote{For $s\leq r'-r$ the intersection volume is actually given by $V_{\cap}=4\pi r^{3}/3$. However, the point
	separation $s$ is always greater than the difference $r'-r$. This
	holds also for the two-bubble case considered below, under the conditions
	imposed by the Heaviside functions in Eq.~(\ref{Pud}) \cite{mm20}.}. 
In brief, the interpretation
of Eq.~(\ref{Pus}) is the following (see \cite{mm20} for details).
The quantity $I(t)$ is related to the probability that the point
$p$ on $S$ has been reached at time $t$ by bubbles nucleated at
times $t''<t$. Such bubbles would have nucleated within the volume
$V=\frac{4\pi}{3}R(t'',t)^{3}$ around $p$. Similarly, $I(t')$ corresponds
to the probability for $p'$ on $S'$, and depends on the volume $V'=\frac{4\pi}{3}R(t'',t')^{3}$.
In order to avoid doubly counting bubbles which would affect both
points, the intersection volume $V_{\cap}$ must be subtracted. Finally,
the bubble containing $p$ and $p'$ would not exist if a previous
bubble were nucleated within a volume $V_{N}=\frac{4\pi}{3}R(t'',t_{N})^{3}$
around its center. The term $I(t_{N})$ takes into account that this
volume must also be subtracted from the calculation.
In the two-bubble case, the joint probability for the points $ p $ and $ p' $ to be uncollided is very similar, 
\begin{align}
P_{uu}^{(d)}= & \,\exp\left[-I(t)-I(t')+I_{\cap}(t,t',s)+I(t_{N})+I(t_{N}^{\prime})\right]\nonumber \\
 & \times\Theta\big(|\mathbf{l}-R\hat{r}|-R(t_{N}^{\prime},t)\big)\:\Theta\big(|\mathbf{l}+R'\hat{r}'|-R(t_{N},t')\big).\label{Pud}
\end{align}
The main difference is that $P_{uu}$ vanishes if the bubbles are so close
that one of the points has been eaten by the other bubble. Thus,
the Heaviside functions take care that the distance from one of the points
to the center of the other bubble be greater than the radius of the
latter at the corresponding time.

Notice that the factors $e^{I(t_{N})}$ and $e^{I(t_{N}^{\prime})}$
in Eqs.~(\ref{Pus}) and (\ref{Pud}) will cancel with the factors
$f_{+}(t_{N})$ and $f_{+}(t_{N}^{\prime})$ in Eqs.~(\ref{Pisfinal-env})
and (\ref{Pidfinal-env}). Thus, using also the surface energy density
(\ref{sigmarhovac}), we obtain 
\begin{align}
\Pi^{(s)}(t,t',\omega)=\, & \frac{\kappa^{2}\rho_{\mathrm{vac}}^{2}}{9}\int_{-\infty}^{t}dt_{N}\Gamma(t_{N})R(t_{N},t)^{3}R(t_{N},t')^{3}\nonumber \\
 & \times\int d\hat{r}\int d\hat{r}'\,e^{-I_{\mathrm{tot}}(t,t',s)}e^{i\omega\hat{n}\cdot\mathbf{s}}\left(\Lambda_{ij,kl}\hat{r}_{i}\hat{r}_{j}\hat{r}'_{k}\hat{r}'_{l}\right),\label{Piscol}
\end{align}
\begin{align}
\Pi^{(d)}(t,t',\omega)= & \,\frac{\kappa^{2}\rho_{\mathrm{vac}}^{2}}{9}\int_{-\infty}^{t}dt_{N}\Gamma(t_{N})R(t_{N},t)^{3}\int_{-\infty}^{t'}dt_{N}^{\prime}\Gamma(t_{N}^{\prime})R(t_{N}^{\prime},t')^{3}\nonumber \\
 & \times\int d^{3}l\int d\hat{r}\int d\hat{r}'\,e^{-I_{\mathrm{tot}}(t,t',s)}e^{i\omega\hat{n}\cdot\mathbf{s}}\left(\Lambda_{ij,kl}\hat{r}_{i}\hat{r}_{j}\hat{r}'_{k}\hat{r}'_{l}\right),\label{Pidcol}
\end{align}
where $I_{\mathrm{tot}}\equiv I(t)+I(t')-I_{\cap}(t,t',s)$. The conditions
imposed by the Heaviside functions (\ref{Pud}) must be taken into
account as limits of integration in Eq.~(\ref{Pidcol}). To simplify
the angular integrations, we shall first average out the direction
of observation $\hat{n}$ (we can do that since the GW spectrum will
not depend on $\hat{n}$). Thus, we make the replacement
\begin{equation}
e^{i\omega\hat{n}\cdot\mathbf{s}}\Lambda_{ij,kl}(\hat{n})\hat{r}_{i}\hat{r}_{j}\hat{r}'_{k}\hat{r}'_{l}\to\frac{1}{4\pi}\int d\hat{n}\,e^{i\omega\hat{n}\cdot\mathbf{s}}\Lambda_{ij,kl}(\hat{n})\hat{r}_{i}\hat{r}_{j}\hat{r}'_{k}\hat{r}'_{l}=\sum_{i=0}^{2}C_{i}\frac{j_{i}(\omega s)}{(\omega s)^{i}}.\label{integn}
\end{equation}
(the calculation is done in appendix \ref{angint}). Here, $j_{i}$
are the spherical Bessel functions 
\begin{equation}
j_{0}(x)=\frac{\sin x}{x},\quad j_{1}(x)=\frac{\sin x-x\cos x}{x^{2}},\quad j_{2}(x)=\frac{(3-x^{2})\sin x-3x\cos x}{x^{3}},\label{SBessel}
\end{equation}
and the coefficients $C_{i}$ are given by 
\begin{align}
C_{0} & =-\frac{1}{2}+\frac{\left(\hat{r}\cdot\hat{s}\right)^{2}}{2}+\frac{\left(\hat{r}'\cdot\hat{s}\right)^{2}}{2}+\frac{\left(\hat{r}\cdot\hat{s}\right)^{2}\left(\hat{r}'\cdot\hat{s}\right)^{2}}{2}-2\left(\hat{r}\cdot\hat{s}\right)\left(\hat{r}'\cdot\hat{s}\right)\left(\hat{r}\cdot\hat{r}'\right)+\left(\hat{r}\cdot\hat{r}'\right)^{2}, 
\label{C0}
\\
C_{1} & =1-\left(\hat{r}\cdot\hat{s}\right)^{2}-\left(\hat{r}'\cdot\hat{s}\right)^{2}-5\left(\hat{r}\cdot\hat{s}\right)^{2}\left(\hat{r}'\cdot\hat{s}\right)^{2}+8\left(\hat{r}\cdot\hat{s}\right)\left(\hat{r}'\cdot\hat{s}\right)\left(\hat{r}\cdot\hat{r}'\right)-2\left(\hat{r}\cdot\hat{r}'\right)^{2}, \label{ci} 
\\
C_{2} & =\frac{1-{5}(\hat{r}\cdot\hat{s})^{2}-{5}(\hat{r}'\cdot\hat{s})^{2}+{35}(\hat{r}\cdot\hat{s})^{2}(\hat{r}'\cdot\hat{s})^{2}}{2}
-10(\hat{r}\cdot\hat{s})(\hat{r}'\cdot\hat{s})(\hat{r}\cdot\hat{r}')+(\hat{r}\cdot\hat{r}')^{2}. 
\label{C2}
\end{align}

For the single-bubble contribution (\ref{Piscol}) we have $\mathbf{s}=R'\hat{r}'-R\hat{r}$,
so all the terms in (\ref{C0})-(\ref{C2}) can be written in terms of a single
cosine, namely, $c\equiv\hat{r}\cdot\hat{r}'$, which, furthermore,
is related to $s$ through
\begin{equation}
s^{2}=R^{2}+R^{\prime2}-2RR'c.\label{s}
\end{equation}
We thus obtain (see appendix \ref{angint})
\begin{equation}
\frac{1}{4\pi}\int d\hat{n}\,e^{i\omega\hat{n}\cdot\mathbf{s}}\left(\Lambda_{ij,kl}\hat{r}_{i}\hat{r}_{j}\hat{r}'_{k}\hat{r}'_{l}\right)=\sum_{i=0}^{2}\frac{P_{i}(s,R,R')}{32R^{2}R^{\prime2}s^{4}}\frac{j_{i}(\omega s)}{(\omega s)^{i}},\label{intns}
\end{equation}
where the $P_{i}$ are homogeneous polynomials of degree 8, which
have simpler expressions in terms of $R_{+}=R'+R$ and $R_{-}=R'-R=R(t,t')$,
\begin{align}
P_{0}(R_{+},R_{-},s)= & \,(s^{2}-R_{-}^{2})^{2}(s^{2}-R_{+}^{2})^{2},\label{P0}\\
P_{1}(R_{+},R_{-},s)= & \,2(s^{2}-R_{-}^{2})(s^{2}-R_{+}^{2})\left[3s^{4}+s^{2}(R_{-}^{2}+R_{+}^{2})-5R_{-}^{2}R_{+}^{2}\right],\\
P_{2}(R_{+},R_{-},s)= & \,3s^{8}+2s^{6}(R_{-}^{2}+R_{+}^{2})+3s^{4}(R_{-}^{4}+4R_{-}^{2}R_{+}^{2}+R_{+}^{4})\label{P2}\\
 & -30s^{2}R_{-}^{2}R_{+}^{2}(R_{-}^{2}+R_{+}^{2})+35R_{-}^{4}R_{+}^{4}.\nonumber 
\end{align}
 Inserting in Eq.~(\ref{Piscol}) and using $\int d\hat{r}\int d\hat{r}'=8\pi^{2}\int_{R_{-}}^{R_{+}}\frac{sds}{RR'}$,
we obtain 
\begin{equation}
\Pi^{(s)}(t,t',\omega)=\frac{\pi^{2}}{4}\frac{\kappa^{2}\rho_{\mathrm{vac}}^{2}}{9}\int_{-\infty}^{t}dt_{N}\Gamma(t_{N})\int_{R_{-}}^{R_{+}}\frac{ds}{s^{3}}\,\sum_{i=0}^{2}P_{i}\frac{j_{i}(\omega s)}{(\omega s)^{i}}e^{-I_{\mathrm{tot}}(t,t',s)}.\label{Piscolfinal}
\end{equation}

For the two-bubble contribution (\ref{Pidcol}), the vector $\mathbf{s}=\mathbf{l}+R'\hat{r}'-R\hat{r}$
is independent of $\hat{r}$ and $\hat{r}'$. Changing the variable
of integration $\mathbf{l}$ to $\mathbf{s}$ and using the result
(\ref{integn}), the second line in (\ref{Pidcol}) becomes 
\begin{equation}
\sum_{i=0}^{2}\int s^{2}ds\,e^{-I_{\mathrm{tot}}(t,t',s)}\frac{j_{i}(\omega s)}{(\omega s)^{i}}\int d\hat{s}\int d\hat{r}\int d\hat{r}'\,C_{i}.\label{bracketd}
\end{equation}
The angular integrations with respect to $\hat{r}$ and $\hat{r}'$
are restrained by the Heaviside functions of Eq.~(\ref{Pud}), which
give conditions on the cosines
\begin{align}
\hat{r}\cdot\hat{s} & \geq-\frac{s^{2}+R^{2}-(R+R_{-})^{2}}{2sR}\equiv-c_{M},\label{cM}\\
\hat{r}'\cdot\hat{s} & \leq\frac{s^{2}+R^{\prime2}-(R'-R_{-})^{2}}{2sR'}\equiv c_{M}^{\prime},\label{cMp}
\end{align}
where we have defined $R_{-}=R(t,t')=\int_{t}^{t'}v_{w}(t'')dt''$
(notice that $R_{-}\neq R'-R$ in this case\footnote{While in the single-bubble case we have $R_{-}=R'-R=R(t,t')$, in
the two-bubble case we have $R'-R\neq R(t,t')$ since the nucleation
times are different in $R=R(t_{N},t)$ and $R'=R(t_{N}^{\prime},t')$.
In this case, the quantity $R'-R$ depends on the nucleation times,
while $R_{-}\equiv R(t,t')$ does not.}). As shown in appendix \ref{angint}, the angular integrals vanish
whenever $c_{M}$ or $c'_{M}$ is outside of the interval
$[-1,+1]$.\footnote{For instance, when $-c_{M}\geq1$, the integral vanishes because there
is no allowed range for the cosine $\hat{r}\cdot\hat{s}$. Physically,
this situation occurs when the bubble of radius $R$ is completely
inside the bubble of radius $R'$ in Fig.~\ref{figdist}. On the
other hand, when $-c_{M}\leq-1$, the whole range $-1\leq \hat{r}\cdot\hat{s}\leq1$
is allowed. This situation occurs when the separation between the
two bubbles is such that they do not overlap, and the integral vanishes due to spherical
symmetry.} If the values of $c_{M}$ and $c'_{M}$ are both in the interval
$[-1,+1]$, the integrals of $C_{0}$ and $C_{1}$ still vanish, while
that of $C_{2}$ gives (see appendix \ref{angint} for details)
\begin{equation}
\int d\hat{s}\int d\hat{r}\int d\hat{r}'\,C_{2}=16\pi^{3}(c_{M}-c_{M}^{3})(c_{M}^{\prime}-c_{M}^{\prime3}).\label{cmcmp}
\end{equation}
The conditions $-1\leq c{}_{M},c'_{M}\leq1$ are equivalent to 
\begin{equation}
s\geq R_{-},\:s\leq2R'-R_{-},\:s\leq2R+R_{-}.\label{limits}
\end{equation}
The latter two imply $s\leq\min\{2R'-R_{-},2R+R_{-}\}=R_{+}-|R(t_{N},t_{N}')|$.
Inserting the result (\ref{cmcmp}) in Eq.~(\ref{bracketd}), with
$c_{M}$ and $c_{M}'$ given by (\ref{cM})-(\ref{cMp}), using the
conditions (\ref{limits}), and finally inserting in Eq.~(\ref{Pidcol}),
we obtain 
\begin{align}
\Pi^{(d)}(t,t',\omega)= & \frac{\pi^{3}}{4}\frac{\kappa^{2}\rho_{\mathrm{vac}}^{2}}{9}\int_{-\infty}^{t}dt_{N}\Gamma(t_{N})\int_{-\infty}^{t'}dt_{N}^{\prime}\Gamma(t_{N}^{\prime})\nonumber \\
 & \times\int_{R_{-}}^{R_{+}-|R(t_{N},t_{N}')|}\frac{ds}{s^{4}}\,e^{-I_{\mathrm{tot}}(t,t',s)}\frac{j_{2}(\omega s)}{(\omega s)^{2}}Q_{+}(s,R,R_{-})Q_{-}(s,R',R_{-}).\label{Pidcolfinal}
\end{align}
where the functions $Q_{\pm}$ are given by
\begin{align}
Q_{+} & (s,R,R_{-})=(s^{2}-R_{-}^{2})\left[(2R+R_{-})^{2}-s^{2}\right]\left[s^{2}-R_{-}(2R+R_{-})\right],\label{Qmas}\\
Q_{-} & (s,R',R_{-})=(s^{2}-R_{-}^{2})\left[(2R'-R_{-})^{2}-s^{2}\right]\left[s^{2}+R_{-}(2R'-R_{-})\right]\label{Qmenos}
\end{align}

It can be seen that this result coincides with that of Ref.~\cite{jt17} for the case of an exponential nucleation rate and a constant wall velocity. 
We shall consider this case as well as other specific models elsewhere \cite{mm21b}. 
Our general expressions 
also seem to be in agreement with the expressions given in the different appendices of Ref.~\cite{jt19}. 
In any case, 
as shown in Appendix \ref{approaches}, our method is equivalent to previous approaches.

\section{The bulk flow model}
\label{bfm}

The bulk flow model \cite{jt19,k18} adapts the envelope approximation
to take into account the fact that the bulk fluid motions do not disappear
when bubbles collide. In this model, the energy is concentrated in
thin fluid shells which form complete (overlapping) spheres. In each
spherical shell, the surface energy density is not homogeneous, since
the fluid is in contact with a bubble wall only in the uncollided
regions of the bubble. The uncollided wall transfers
an amount of energy  to the fluid which is proportional to the bubble volume, and
on the other hand
the surface energy density decreases with the surface area,
so it is given by Eq.~(\ref{sigmarhovac}). In a collided region,
in contrast, the wall has disappeared and the fluid does not receive energy anymore. 
The fluid shell is assumed to move with the same velocity as the bubble
wall (for consistency, the latter must be a constant). 
The energy density in that
part of the fluid shell will thus be proportional to $R_{c}^{2}/R^{2}$,
where $R=v(t-t_{N})$ is the bubble radius and $R_{c}=v(t_{c}-t_{N})$
its value at the moment of collision $t_{c}$. The latter depends
on the specific point on the bubble surface, which we represent by
the unit vector $\hat{r}$. The stress-energy tensor for a bubble
at time $t$ is similar to that of the envelope approximation, 
Eqs.~(\ref{Tijenv})-(\ref{indicator}). In the present
case we have 
\begin{equation}
T_{ij}=\hat{r}_{i}\hat{r}_{j}\,\delta(r-R)\,\sigma(\hat{r}),
\end{equation}
with 
\begin{equation}
\sigma(\hat{r})=\begin{cases}
\sigma_{u}(t) & \mathrm{for}\:t_{c}(\hat{r})>t,\\
\sigma_{c}(t_{c},t) & \mathrm{for}\:t_{c}(\hat{r})<t,
\end{cases}\label{sigmabf}
\end{equation}
where $\sigma_{u}=(\kappa\rho_{\mathrm{vac}}/3)R$ is the surface
energy density at an uncollided point, while $\sigma_{c}$ is the
surface energy density at a collided point. The latter is given by
\begin{equation}
\sigma_{c}=\frac{\kappa\rho_{\mathrm{vac}}}{3}R_{c}\times\frac{R_{c}^{2}}{R^{2}}\times D(t_{c},t).\label{sigmac}
\end{equation}
The first factor in (\ref{sigmac}) is the value of $\sigma_{u}$
at the the collision time of the given point. The second factor takes
into account the scaling as $R^{-2}$, and the third factor is a damping
function introduced to account for an additional energy loss \cite{jt19}.
We thus have 
\begin{equation}
\sigma(\hat{r})=\sigma_{u}(t)\begin{cases}
1 & \mathrm{for}\:t_{c}(\hat{r})>t,\\
\frac{(t_{c}-t_{N})^{3}}{(t-t_{N})^{3}}D(t_{c},t) & \mathrm{for}\:t_{c}(\hat{r})<t,
\end{cases}\label{sigmabf-1}
\end{equation}

Repeating the steps of Sec.~\ref{bbcoll}, we readily arrive at expressions
for $\Pi^{(s)}$ and $\Pi^{(d)}$ similar to Eqs.~(\ref{Pisfinal-env})
and (\ref{Pidfinal-env}), 
\begin{align}
\Pi^{(s)}(t,t',\omega)= & \,\int_{-\infty}^{t}dt_{N}\Gamma(t_{N})f_{+}(t_{N})R^{2}R^{\prime2}\int d\hat{r}\int d\hat{r}'\,e^{i\omega\hat{n}\cdot\mathbf{s}}\Lambda_{ij,kl}\hat{r}_{i}\hat{r}_{j}\hat{r}'_{k}\hat{r}'_{l}\left\langle \sigma\sigma'\right\rangle ,\label{Pisfinal-bf}\\
\Pi^{(d)}(t,t',\omega)= & \,\int_{-\infty}^{t}dt_{N}\Gamma(t_{N})f_{+}(t_{N})R^{2}\int_{-\infty}^{t'}dt_{N}^{\prime}\Gamma(t_{N}^{\prime})f_{+}(t_{N}^{\prime})R^{\prime2}\nonumber \\
 & \,\times\int d^{3}l\int d\hat{r}\int d\hat{r}'\,e^{i\omega\hat{n}\cdot\mathbf{s}}\Lambda_{ij,kl}\hat{r}_{i}\hat{r}_{j}\hat{r}'_{k}\hat{r}'_{l}\left\langle \sigma\sigma'\right\rangle ,\label{Pidfinal-bf}
\end{align}
where $\sigma'$ is the surface energy density at angular position
$\hat{r}'$ on the bubble of radius $R'=v(t'-t_{N}')$ at time $t'$ 
(with $t_{N}'=t_{N}$ for the single-bubble case).

We must evaluate the average $\left\langle \sigma\sigma'\right\rangle $
at fixed times $t,t'$ and fixed directions $\hat{r},\hat{r}'$. The
random variables here are the collision times $t_{c}(\hat{r}),t_{c}(\hat{r}')$
of each point. Notice that 
a given point on a
bubble surface will eventually collide, and the probability for the collision time
$t_{c}$ does not depend on the time $t$
(before or after $t_{c}$) at which we evaluate the average. This
probability does depend on the knowledge of another point 
(on the same or a different bubble) being collided or not.  
Hence, we need to calculate the joint probability $dP_{cc}$
that the collision time of the point at $\hat{r}$ is in the interval
$[t_{c},t_{c}+dt_{c}]$ and the collision time of the point at $\hat{r}'$
is in the interval $[t_{c}^{\prime},t_{c}^{\prime}+dt_{c}^{\prime}]$.
This infinitesimal probability is related to the probability $P_{cc}$
of both points having collided before the times $t_{c}$ and $t_{c}^{\prime}$
through 
\begin{equation}
dP_{cc}=\frac{\partial^{2}P_{cc}}{\partial t_{c}\partial t_{c}^{\prime}}dt_{c}dt_{c}^{\prime}.\label{dpc}
\end{equation}
On the other hand, the probability $P_{cc}$ is related to the probability
$P_{uu}$ that both points are uncollided, which is given by Eq.~(\ref{Pus})
for the single-bubble case and by Eq.~(\ref{Pud}) for the two-bubble
case. We have\footnote{The normalization condition for probability implies that $P_{cc}=1-P_{uu}-P_{uc}-P_{cu}$,
where $P_{cu}$ and $P_{uc}$ are the probabilities that only one
of the points has collided before the corresponding time $t_{c}$
or $t_{c}'$. These are given by $P_{u}(t_{c})-P_{uu}$ and $P_{u}(t_{c}')-P_{uu}$,
where $P_{u}(t)=e^{-I(t)+I(t_{N})}$ is the independent probability
that a point on the bubble surface is uncollided at time $t$.} 
\begin{equation}
P_{cc}=1-P_{u}(t_{c})-P_{u}(t_{c}^{\prime})+P_{uu}(t_{c},t_{c}^{\prime},s_{c}),\label{Pc}
\end{equation}
where $P_{u}(t)$ is the probability that a single point on the bubble
surface is uncollided at time $t$. The first three terms in (\ref{Pc})
vanish when we take the second derivative in (\ref{dpc}), and we
have
\begin{equation}
\left\langle \sigma\sigma'\right\rangle =\int\sigma\sigma'dP_{cc}=\int_{t_{N}}^{\infty}\int_{t_{N}^{\prime}}^{\infty}\sigma\sigma'\frac{\partial^{2}P_{uu}}{\partial t_{c}\partial t_{c}^{\prime}}dt_{c}dt_{c}^{\prime}.
\end{equation}
Integrating by parts and taking into account that $P_{uu}\to0$ for
$t_{c}\to\infty$ or $t_{c}'\to\infty$ (in which case one of the
points is certainly collided) and that $\sigma(t_{c}=t_{N})=\sigma'(t_{c}'=t_{N}')=0$
(corresponding to a collision as soon as the bubble nucleates), we
obtain
\begin{equation}
\left\langle \sigma\sigma'\right\rangle =\int_{t_{N}}^{\infty}\int_{t_{N}^{\prime}}^{\infty}\frac{\partial^{2}(\sigma\sigma')}{\partial t_{c}\partial t_{c}^{\prime}}P_{uu}\,dt_{c}dt_{c}^{\prime}=\frac{\sigma_{u}\sigma_{u}^{\prime}}{R^{\prime3}R^{3}}\int_{t_{N}}^{t}dt_{c}\frac{\partial R_{c}^{3}D}{\partial t_{c}}\int_{t_{N}^{\prime}}^{t'}dt_{c}^{\prime}\frac{\partial R_{c}^{\prime3}D'}{\partial t_{c}^{\prime}}\,P_{uu}\label{sspm}
\end{equation}
where we have used Eqs.~(\ref{sigmabf-1}). From now on, we shall
consider for simplicity the case $D=1$ (free propagation with no
damping). Thus, we have 
\begin{equation}
\left\langle \sigma\sigma'\right\rangle =(\kappa\rho_{\mathrm{vac}})^{2}v^{2}\int_{t_{N}}^{t}dt_{c}\frac{R_{c}^{2}}{R^{2}}\int_{t_{N}^{\prime}}^{t'}dt_{c}^{\prime}\frac{R_{c}^{\prime2}}{R^{\prime2}}\,P_{uu}(t_{c},t_{c}^{\prime},s_{c}).\label{sspm-1}
\end{equation}

Since Eqs.~(\ref{Pisfinal-bf})-(\ref{Pidfinal-bf}) are very similar
to (\ref{Pisfinal-env})-(\ref{Pidfinal-env}), we may proceed like
in Sec.~\ref{bbcoll}. In particular, the average over the direction
of observation $\hat{n}$ introduces the factor (\ref{integn}),
\begin{equation}
\sum_{i=0}^{2}\frac{j_{i}(\omega s)}{(\omega s)^{i}}C_{i}(\hat{r}\cdot\hat{s},\hat{r}'\cdot\hat{s},\hat{r}\cdot\hat{r}'),\label{intns1}
\end{equation}
where the functions $C_{i}$ are defined in Eqs.~(\ref{C0})-(\ref{C2}), and
$\mathbf{s}=\mathbf{l}+R'\hat{r}'-R\hat{r}$ (with $\mathbf{l}=\mathbf{0}$
for the single-bubble case). The dependence on the vector $\mathbf{s}$
comes from the factor $e^{i\omega\hat{n}\cdot\mathbf{s}}$ in (\ref{Pisfinal-bf})-(\ref{Pidfinal-bf}).
The main difference with the envelope approximation is that in that
case we had
\begin{equation}
\left\langle \sigma\sigma'\right\rangle =\sigma_{u}\sigma_{u}'R^{2}R^{\prime2}\left\langle 1_{S}(\hat{r})1_{S'}(\hat{r}')\right\rangle =(\kappa\rho_{\mathrm{vac}}/3)^{2}RR'P_{uu}(t,t',s).
\end{equation}
In the present case this expression must
be replaced with (\ref{sspm-1}), which depends on the vector $\mathbf{s}_{c}=\mathbf{l}+R_{c}'\hat{r}'-R_{c}\hat{r}$
joining the positions of the points at the collision times. The probability
$P_{uu}$ depends in this case on the quantity 
\begin{equation}
I_{\cap}(t_{c},t_{c}',s_{c})=\int_{-\infty}^{\min\{t_{c},t_{c}'\}}dt''\Gamma(t'')V_{\cap}(t'',t_{c},t_{c}',s_{c}).
\end{equation}

For the single-bubble contribution, the distance $s$ is given by
Eq.~(\ref{s}), and $s_{c}$ has a similar expression in terms of
$c=\hat{r}\cdot\hat{r}'$, which is the only independent angular variable
in the integrand of (\ref{Pisfinal-bf}). The variables $s$ and $s_{c}$
are thus related by
\begin{equation}
c=\frac{R^{2}+R^{\prime2}-s^{2}}{2RR'}=\frac{R_{c}^{2}+R_{c}^{\prime2}-s_{c}^{2}}{2R_{c}R_{c}'},
\label{sscs}
\end{equation}
and the angular integrals in (\ref{Pisfinal-bf}) can be written
in several forms,
\begin{equation}
\int d\hat{r}\int d\hat{r}'=8\pi^{2}\int_{-1}^{+1}dc=8\pi^{2}\int_{R_{-}}^{R_{+}}\frac{sds}{RR'}=8\pi^{2}\int_{R_{c-}}^{R_{c+}}\frac{s_{c}ds_{c}}{R_{c}R_{c}'},
\end{equation}
with $R_{c\pm}=R_{c}'\pm R_{c}$. If we write $\hat{r}\cdot\hat{s}$
and $\hat{r}'\cdot\hat{s}$ in terms of $c$, Eq.~(\ref{intns1})
gives the expression (\ref{intns}). Then, if we choose $s$ as the integration
variable, like we did in Sec.~\ref{bbcoll}, we obtain
\begin{align}
\Pi^{(s)}(t,t',\omega)= & \,\frac{\pi^{2}}{4}(\kappa\rho_{\mathrm{vac}})^{2}v^{2}\int_{-\infty}^{t}dt_{N}\Gamma(t_{N})\int_{R_{-}}^{R_{+}}\frac{ds}{s^{3}}\sum_{i=0}^{2}\frac{P_{i}(s,R,R')}{R^{\prime3}R^{3}}\frac{j_{i}(\omega s)}{(\omega s)^{i}}\nonumber \\
 & \times\int_{t_{N}}^{t}dt_{c}R_{c}^{2}\int_{t_{N}}^{t'}dt_{c}^{\prime}R_{c}^{\prime2}\,e^{-I_{\mathrm{tot}}(t_{c},t_{c}',s_{c})}.
 \label{Pisfinal-bf-1}
\end{align}
This expression is equivalent to Eq.~(B.9) of Ref.~\cite{jt19}.

For the two-bubble contribution, the integrations on $\hat{r}$ and
$\hat{r}'$ are restrained by the Heaviside functions of Eq.~(\ref{Pud})
at the times $t_{c}$ and $t_{c}'$, which give conditions like (\ref{cM})-(\ref{cMp})
but for the cosines $\hat{r}\cdot\hat{s}_{c}$ and $\hat{r}'\cdot\hat{s}_{c}$,
\begin{align}
\hat{r}\cdot\hat{s}_{c} & \geq-\frac{s_{c}^{2}+R_{c}^{2}-(R_{c}+R_{c-})^{2}}{2s_{c}R_{c}}\equiv-c_{Mc},\label{cM-1}\\
\hat{r}'\cdot\hat{s}_{c} & \leq\frac{s_{c}^{2}+R_{c}^{\prime2}-(R'_{c}-R_{c-})^{2}}{2s_{c}R_{c}'}\equiv c_{Mc}^{\prime},\label{cMp-1}
\end{align}
where $R_{c-}=R(t_{c},t_{c}')$. In contrast, the quantities $C_{i}$
in (\ref{intns1}) depend on the cosines $\hat{r}\cdot\hat{s}$, $\hat{r}'\cdot\hat{s}$.
The variables at the times $t,t'$ are related to those at the collision
times $t_{c},t_{c}'$ by 
\begin{equation}
\mathbf{l}=\mathbf{s}-R'\hat{r}'+R\hat{r}=\mathbf{s}_{c}-R_{c}'\hat{r}'+R_{c}\hat{r}.\label{lssc}
\end{equation}
Using (\ref{lssc}), we may change the variable of integration $\mathbf{l}$
in Eq.~(\ref{Pidfinal-bf}) either to $\mathbf{s}$ or to $\mathbf{s}_{c}$.
The first choice leads to an expression which is equivalent to Eq.~(B.18)
of Ref.~\cite{jt19}. In this case it is convenient to define angular
variables $c_{r}=\hat{r}\cdot\hat{s}$, $c_{r'}=\hat{r}\cdot\hat{s}$
and the azimuth $\phi_{r'}$ like we did in appendix \ref{angint},
\begin{multline}
\frac{\Pi^{(d)}}{\kappa^{2}\rho_{\mathrm{vac}}^{2}}=8\pi^{2}v^{2}\int_{-\infty}^{t}dt_{N}\Gamma(t_{N})\int_{-\infty}^{t'}dt_{N}^{\prime}\Gamma(t_{N}^{\prime})\int_{t_{N}}^{t}dt_{c}R_{c}^{2}\int_{t_{N}^{\prime}}^{t'}dt_{c}^{\prime}R_{c}^{\prime2}\\
\times\int_{0}^{\infty}s^{2}ds\sum_{i=0}^{2}\frac{j_{i}(\omega s)}{(\omega s)^{i}}\int_{-1}^{1}dc_{r}\int_{-1}^{1}dc_{r'}\int_{0}^{2\pi}d\phi_{r'}\,C_{i}(c_{r},c_{r'},\hat{r}\cdot\hat{r}')e^{-I_{\mathrm{tot}}(t_{c},t_{c}',s_{c})}\\
\times\Theta(\hat{r}\cdot\hat{s}_{c}+c_{Mc})\Theta(c_{Mc}'-\hat{r}'\cdot\hat{s}_{c}).
\end{multline}
In the envelope approximation we made analytically the integrals $\int d\phi_{r'}C_{i}$.
However, in this case the variable $s_{c}$ also depends on $\phi_{r'}$.

Notice also that the Heaviside functions imposing the conditions (\ref{cM-1})-(\ref{cMp-1})
upon the integration domain are non-trivial in terms of the variables
$c_{r},c_{r'}$. A somewhat simpler expression is obtained if we use
$\mathbf{s}_{c}$ instead of $\mathbf{s}$ as integration variable,
since these restrictions can be implemented explicitly in the limits
of integration. We proceed like we did in appendix \ref{angint},
but we define angular variables with respect to $\mathbf{s}_{c}$
instead of $\mathbf{s}$, i.e., $c_{rc}=\hat{r}\cdot\hat{s}_{c}$,
$c_{r'c}=\hat{r}'\cdot\hat{s}_{c}$, and the azimuth $\phi_{r'c}$
of $\hat{r}'$. We obtain
\begin{multline}
\frac{\Pi^{(d)}}{\kappa^{2}\rho_{\mathrm{vac}}^{2}}=8\pi^{2}v^{2}\int_{-\infty}^{t}\negmedspace dt_{N}\Gamma(t_{N})\int_{-\infty}^{t'}\negmedspace dt_{N}^{\prime}\Gamma(t_{N}^{\prime})\int_{t_{N}}^{t}\negmedspace dt_{c}R_{c}^{2}\int_{t_{N}^{\prime}}^{t'}\negmedspace dt_{c}^{\prime}R_{c}^{\prime2}\int_{0}^{\infty}\negmedspace ds_{c}s_{c}^{2}\,e^{-I_{\mathrm{tot}}(t_{c},t_{c}',s_{c})}\\
\times\int_{\max\{-c_{Mc},-1\}}^{1}\negthickspace dc_{rc}\int_{-1}^{\min\{c_{Mc}^{\prime},1\}}\negthickspace dc_{r'c}\int_{0}^{2\pi}d\phi_{r'c}\sum_{i=0}^{2}\frac{j_{i}(\omega s)}{(\omega s)^{i}}C_{i}(\hat{r}\cdot\hat{s},\hat{r}'\cdot\hat{s},\hat{r}\cdot\hat{r}').
\label{Pidfinal-bf-1}
\end{multline}
Finally, the relation between the variables with and without the index
$c$ is obtained form Eq.~(\ref{lssc}),
\begin{equation}
s^{2}=s_{c}^{2}+\Delta R^{\prime2}+\Delta R^{2}+2s_{c}\Delta R'(\hat{r}'\cdot\hat{s}_{c})-2s_{c}\Delta R(\hat{r}\cdot\hat{s}_{c})-2\Delta R'\Delta R(\hat{r}\cdot\hat{r}'),
\end{equation}
\begin{equation}
\hat{r}\cdot\hat{s}=\frac{s_{c}\,\hat{r}\cdot\hat{s}_{c}+\Delta R'(\hat{r}\cdot\hat{r}')-\Delta R}{s},\quad\hat{r}'\cdot\hat{s}=\frac{s_{c}\,\hat{r}'\cdot\hat{s}_{c}+\Delta R'-\Delta R(\hat{r}\cdot\hat{r}')}{s},
\end{equation}
where $\Delta R=R-R_{c}=v(t-t_{c})$, $\Delta R'=R'-R'_{c}=v(t'-t_{c}')$,
and 
\begin{equation}
\hat{r}\cdot\hat{r}'=s_{rc}s_{r'c}\cos\phi_{r'c}+c_{rc}c_{r'c},
\end{equation}
with $s_{rc}^{2}=1-c_{rc}^{2}$, and $s_{r'c}^{2}=1-c_{r'c}^{2}$.

We shall apply these expressions to specific phase transition models elsewhere. 
Here we shall focus on the case of non-spherical bubbles.

\section{Wall deformations}
\label{wdef}

We shall now adapt the derivation of $\Pi^{(s)}$ and $\Pi^{(d)}$
of Sec.~\ref{bbcoll} to the case of bubble walls which
depart from the spherical shape. 

\subsection{A general wall shape}

In general, the wall surface can
be parametrized by a function of two variables. For instance, 
$\mathbf{r}=\mathbf{r}_{w}(\hat{r})={r}_{w}(\hat{r})\hat r$
in terms of the angles of the direction $\hat{r}$ from the nucleation center.
It is a good approximation to consider a field profile of the
form \cite{safran}
\begin{equation}
\phi(\mathbf{r})=\phi_{0}\big((\mathbf{r}-\mathbf{r}_{w})\cdot\hat{n}_{S}\big),
\end{equation}
where $\phi_{0}(x)$ is the bubble profile for the one-dimensional
problem and $\hat{n}_{S}$ is the normal to the surface. The function
$\phi_{0}(x)$ varies only in a certain interval of length $l_{w}$
(the wall width) around $x=0$ and takes approximately constant values
outside this interval. If the implicit form of the surface is defined
by an equation $F(\mathbf{r})=0$, we have $\hat{n}_{S}=\nabla F/|\nabla F|$.
For instance, for a spherical bubble, the function $F$ can be taken
as $F(\mathbf{r})=r-R$ (with $r=\sqrt{\mathbf{r}^{2}}$). This gives
$\hat{n}_{S}=\nabla F=\hat{r}$ and $\phi(\mathbf{r})=\phi_{0}(r-R)$,
which is the form assumed in Sec.~\ref{bbcoll}. For non-spherical
bubbles, we write the wall position as $r_{w}=R+\zeta$. Then,
for $F=r-R-\zeta$, we have $\nabla F=\hat{r}-\nabla\zeta$. Since
$\zeta$ only varies in the angular directions, we have $\nabla\zeta\cdot\hat{r}=0$
and $|\nabla F|=\sqrt{1+(\nabla\zeta)^{2}}$. We thus obtain
\begin{equation}
\phi(r,\hat{r},t)=\phi_{0}\left(\frac{r-R-\zeta}{\sqrt{1+(\nabla\zeta)^{2}}}\right),\label{ficomp}
\end{equation}
where the time dependence is implicit in the undeformed wall position
$R(t_{N},t)$ as well as in the deformation $\zeta(\hat{r},t)$.

Using Eq.~(\ref{ficomp}) in $T_{ij}=\partial_{i}\phi\partial_{j}\phi$
we obtain 
\begin{equation}
T_{ij}=\left[\phi_{0}^{\prime}\left(\frac{r-R-\zeta}{\sqrt{1+(\nabla\zeta)^{2}}}\right)\right]^{2}\frac{\left(\hat{r}_{i}-\partial_{i}\zeta\right)\left(\hat{r}_{j}-\partial_{j}\zeta\right)+\mathcal{O}(r-r_{w})}{1+\left(\nabla\zeta\right)^{2}}.\label{Tijfi0}
\end{equation}
We do not write down the terms of order $r-r_{w}$, which vanish in
the thin wall approximation. Like in Sec.~\ref{bbcoll}, we make the approximation
$[\phi_{0}^{\prime}(x)]^{2}=\sigma_{0}\delta(x)$, 
and the first factor in (\ref{Tijfi0}) 
becomes\footnote{Notice that Eq.~(\ref{deltaS}) may be written as $\sigma_{0}\delta(F)|\nabla F|=\sigma_{0}\delta_{S}$,
where $\delta_{S}$ is the surface delta, such that $\int\delta_{S}\varphi d^{3}x=\int_{S}\varphi dS$
for any test function $\varphi$.} 
\begin{equation}
\sigma_{0}\delta\left(\frac{r-R-\zeta}{\sqrt{1+(\nabla\zeta)^{2}}}\right)=\sigma_{0}\delta\left(r-R-\zeta\right)\sqrt{1+(\nabla\zeta)^{2}}.\label{deltaS}
\end{equation}
Like in the envelope approximation, we shall assume that the wall
profile $\phi_{0}$ just disappears in the intersections with other
bubbles. Thus, we have\footnote{To verify this result, let us consider the energy density of the scalar
field in the static case, $T_{00}=\frac{1}{2}(\nabla\phi)^{2}+V(\phi)=(\nabla\phi)^{2}$
(the last equality follows from the field equation inside the wall).
Therefore, we have $T_{00}=\partial_{i}\phi\partial_{i}\phi=T_{ii}$.
According to the result (\ref{Tijzeta}), we thus have $T_{00}=\sigma_{0}\delta(F)|\nabla F|1_{S}=\sigma_{0}\delta_{S}1_{S}$,
as expected.}
\begin{equation}
T_{ij}=\frac{\left(\hat{r}_{i}-\partial_{i}\zeta\right)\left(\hat{r}_{j}-\partial_{j}\zeta\right)}{\sqrt{1+(\nabla\zeta)^{2}}}\,\sigma_{0}\delta(r-R-\zeta)\,1_{S}(\hat{r}),\label{Tijzeta}
\end{equation}
and its spatial Fourier transform, defined by Eq.~(\ref{Ttilden}),
is given by 
\begin{equation}
\tilde{T}_{ij}(t,\omega\hat{n})=\sigma_{0}R^{2}\int d\hat{r}e^{-i\omega R\hat{n}\cdot\hat{r}}t_{ij}1_{S}(\hat{r}),\label{Ttilij}
\end{equation}
where 
\begin{equation}
t_{ij}=\frac{\left(1+\zeta/R\right)^{2}}{\sqrt{1+(\nabla\zeta)^{2}}}e^{-i\omega\zeta\hat{n}\cdot\hat{r}}(\hat{r}_{i}-\partial_{i}\zeta)(\hat{r}_{j}-\partial_{j}\zeta).
\label{tij}
\end{equation}
We shall again replace the parameter $\sigma_{0}$ by a varying quantity $\sigma$,
so that we allow for the possibility that the surface energy density
is not a constant. Inserting Eq.~(\ref{Ttilij}) into Eqs.~(\ref{Pis})
and (\ref{Pid}), we obtain expressions similar to Eqs.~(\ref{Pisfinal-env})-(\ref{Pidfinal-env}),
\begin{equation}
\Pi^{(s)}=\int_{-\infty}^{t}\negmedspace dt_{N}\Gamma(t_{N})f_{+}(t_{N})\sigma R^{2}\sigma'R^{\prime2}\int d\hat{r}\int d\hat{r}'e^{i\omega\hat{n}\cdot\mathbf{s}}\Lambda_{ij,kl}\left\langle t_{ij}t_{kl}^{\prime*}\,1_{S}(\hat{r})1_{S'}(\hat{r}')\right\rangle ,\label{Pisfinal}
\end{equation}
\begin{align}
\Pi^{(d)}= & \,\int_{-\infty}^{t}\negmedspace dt_{N}\Gamma(t_{N})f_{+}(t_{N})\sigma R^{2}\int_{-\infty}^{t'}\negmedspace dt_{N}^{\prime}\Gamma(t_{N}^{\prime})f_{+}(t_{N}^{\prime})\sigma'R^{\prime2}\nonumber \\
 & \times\int d^{3}l\int d\hat{r}\int d\hat{r}'\,e^{i\omega\hat{n}\cdot\mathbf{s}}\Lambda_{ij,kl}\left\langle t_{ij}t_{kl}^{\prime*}\,1_{S}(\hat{r})1_{S'}(\hat{r}')\right\rangle .\label{Pidfinal}
\end{align}
Like in previous sections, a prime in a quantity means that it is
evaluated at $t'$ and $\hat{r}'$, and the definition of the vector
$\mathbf{s}$ is the same. Thus, $t'_{kl}$ is given by Eq.~(\ref{tij})
with $R$ replaced by $R'$, $\hat{r}$ by $\hat{r}'$, and $\zeta$
by $\zeta'\equiv\zeta(\hat{r}',t')$.

The ensemble average in Eqs.~(\ref{Pisfinal})-(\ref{Pidfinal})
involves a statistical treatment of the deformation $\zeta$ as well
as that of the uncollided surfaces $S$ and $S'$. In many cases,
the dynamics of the wall deformation $\zeta$ will be independent
of that of the uncollided surfaces. We expect this to be the case
when the deformations originate from
random perturbations. In such a case, the average separates, 
\begin{equation}
\left\langle t_{ij}t_{kl}^{\prime*}\,1_{S}(\hat{r})1_{S'}(\hat{r}')\right\rangle =\left\langle t_{ij}t_{kl}^{\prime*}\right\rangle \left\langle 1_{S}(\hat{r})1_{S'}(\hat{r}')\right\rangle =
\left\langle t_{ij}t_{kl}^{\prime*}\right\rangle P_{uu}(\hat{r},\hat{r}')
\end{equation}
where, assuming that the phase transition dynamics does not change significantly for deformed bubbles, $P_{uu}$ is given by Eqs.~(\ref{Pus}) and (\ref{Pud}). 
We thus obtain
\begin{equation}
\Pi^{(s)} =
\int_{-\infty}^{t}dt_{N}\Gamma(t_{N})\sigma R^{2}\sigma'R^{\prime2}\int d\hat{r}\int d\hat{r}'\,e^{i\omega\hat{n}\cdot\mathbf{s}} e^{-I_{\mathrm{tot}}(t,t',s)}
\Lambda_{ij,kl}\left\langle t_{ij}t_{kl}^{\prime*}\right\rangle , \label{Pissepar}
\end{equation}
with $\mathbf{s}=R'\hat{r}'-R\hat{r}$, while for the two-bubble contribution  the vector $ \mathbf{s} $ 
is independent of $\hat{r}$ and $\hat{r}'$ and we may change the variable of integration $\mathbf{l}$ to $\mathbf{s}$ like we did in Sec.~\ref{bbcoll}.
It is also reasonable to assume that there is no correlation between
deformations produced in different bubbles, so in this case we have 
\begin{equation}
	\Pi^{(d)}= \int_{-\infty}^{t}\negmedspace dt_{N}\Gamma(t_{N}) \negmedspace 	\int_{-\infty}^{t'}\negmedspace dt_{N}^{\prime}
	\Gamma(t_{N}^{\prime})\sigma R^{2}\sigma'R^{\prime2} \negmedspace  \int \negmedspace d^{3}s \,  e^{i\omega\hat{n}\cdot\mathbf{s}}  e^{-I_{\mathrm{tot}}} 
	\negmedspace \int \negmedspace d\hat{r} 	 \int \negmedspace d\hat{r}'\Lambda_{ij,kl}
	 \langle t_{ij}\rangle\langle t_{kl}^{\prime*}\rangle ,
	\label{Pidsepar}
\end{equation}
where the angular integrals are bound by the conditions (\ref{cM})-(\ref{cMp}).

\subsection{Statistical properties of the deformations}

The averages which remain to be evaluated in Eqs.~(\ref{Pissepar})-(\ref{Pidsepar}) depend on the statistics of $ \zeta $.
Assuming for simplicity a Gaussian probability distribution for the deformations, we only need to 
calculate the quantity $\langle\zeta\zeta'\rangle$ and its derivatives. 
For points on different bubbles we assume that $ \langle \zeta\zeta' \rangle =\langle \zeta\rangle\langle \zeta' \rangle =0$, so we consider two points on a given bubble.
Although the calculation cannot be done without considering a specific deformation mechanism, 
the general properties of these quantities are determined by the symmetry of the problem,
since we consider deformations from a spherical shape.

The expectation value $\langle\zeta\zeta'\rangle$ depends on the times $t,t'$, as well as on the
angular separation $\theta$ between $\hat{r}$ and $\hat{r}'$. We
thus write
\begin{equation}
	\langle\zeta\zeta'\rangle\equiv\alpha(c,t,t'),\label{zetazetap}
\end{equation}
where $c\equiv \cos\theta\equiv \hat{r}\cdot\hat{r}'$,
and we have\footnote{Using $\hat{r}_{i}=x_{i}/r$,
	$\partial(\hat{r}\cdot\hat{r}')/\partial x_{i}=r^{-1}\left[\hat{r}'-(\hat{r}'\cdot\hat{r})\hat{r}\right]_{i}$.}
\begin{equation}
	\langle\zeta'\partial_{i}\zeta\rangle=
	\frac{\partial}{\partial x_{i}} \langle\zeta'\zeta\rangle=
	\frac{\hat{r}_{i}'-c\hat{r}_{i}}{R}\frac{\partial\alpha}{\partial c}.
	\label{dzetazetap}
\end{equation}
Similarly,
\begin{equation}
	\langle\zeta\partial_{k}\zeta'\rangle=
	\frac{\partial}{\partial x_{k}'}\langle\zeta\zeta'\rangle=
	\frac{\hat{r}_{k}-c\hat{r}_{k}'}{R'}\frac{\partial\alpha}{\partial c} 
	\label{zetadzetap}
\end{equation}
and
\begin{equation}
	\langle\partial_{i}\zeta\partial_{k}\zeta'\rangle = \frac{\partial}{\partial x_{k}'}\frac{\partial}{\partial x_{i}}\langle\zeta\zeta'\rangle=
	\frac{\delta_{ik}-\hat{r}_{i}\hat{r}_{k}-\hat{r}_{i}'\hat{r}_{k}'+c\hat{r}_{i}\hat{r}_{k}'}{RR'}\frac{\partial\alpha}{\partial c}+\frac{\left(\hat{r}_{i}'-c\hat{r}_{i}\right)\left(\hat{r}_{k}-c\hat{r}_{k}'\right)}{RR'}\frac{\partial^{2}\alpha}{\partial c^{2}}.
	\label{dzetadzetap}
\end{equation}
For $\hat{r}'\to\hat{r}$, we have $ c\to 1 $ and  Eqs.~(\ref{dzetazetap})-(\ref{zetadzetap}) become 
\begin{equation}
	\langle\zeta'\partial_{i}\zeta\rangle=\langle\zeta\partial_{k}\zeta'\rangle=0 .
	\label{zetadzetap0}
\end{equation}
This was expected by symmetry, since $ \nabla \zeta\perp \hat r $. Similarly, in this case Eq.(\ref{dzetadzetap}) becomes
\begin{equation}
	\langle\partial_{i}\zeta\partial_{k}\zeta'\rangle = \frac{\delta_{ik}-\hat{r}_{i}\hat{r}_{k}}{RR'}\frac{\partial\alpha}{\partial c}.
\end{equation}
which is also expected by symmetry.
In particular, at a given point on a bubble wall, i.e., for $\hat r ' = \hat{r}$ and $ t'=t $, we have 	
\begin{equation}
	\langle\zeta^{2}\rangle=
	\alpha_{0}(t),\quad\langle\zeta\partial_{i}\zeta\rangle=0,\quad
	\langle\partial_{i}\zeta\partial_{j}\zeta\rangle=\beta_{0}(t)\left(\delta_{ij}-\hat{r}_{i}\hat{r}_{j}\right),
	\label{correlssamepoint}
\end{equation}
and $\langle(\nabla\zeta)^{2}\rangle=2\beta_{0}(t)$, where 
$\alpha_{0}(t)\equiv\alpha(1,t,t)$ and $\beta_{0}(t)\equiv R^{-2}\partial_c\alpha(1,t,t)$.

It is also useful to consider the expansion of the wall displacement in spherical harmonics, 
\begin{equation}
	\zeta(\hat{r},t)=\sum_{l=0}^{\infty}\sum_{m=-l}^{+l}D_{lm}(t)Y_{l}^{m}(\hat{r}).
\end{equation}
By symmetry, the correlator for the amplitude is of the form
\begin{equation}
	\left\langle D_{lm}(t)D_{l'm'}(t')^{*}\right\rangle \equiv\delta_{ll'}\delta_{mm'}\,|D|_{l}^{2}(t,t'),
	\label{displacement}
\end{equation}
which yields (using the addition theorem of the spherical harmonics)
\begin{equation}
	\alpha(c,t,t')=\left\langle \zeta(\hat{r},t)\zeta(\hat{r}',t')\right\rangle  = \sum_{l=0}^{\infty}\frac{2l+1}{4\pi}|D|_{l}^{2}(t,t')P_{l}(c),
	\label{alfa}
\end{equation}
where $ P_l $ are the Legendre polynomials.

\subsection{Small deformations}

If we expand the quantity $t_{ij}$ in powers of $\zeta$ and $ \partial_i\zeta $, we obtain $ t_{ij}=t_{ij}^{(0)}+t_{ij}^{(1)}+\cdots $, with
\begin{align}
	t_{ij}^{(0)} & =\hat{r}_{i}\hat{r}_{j}e^{-i\omega\zeta\hat{n}\cdot\hat{r}},\label{tij0-1}\\
	t_{ij}^{(1)} & =\left[2\hat{r}_{i}\hat{r}_{j}\frac{\zeta}{R}-\hat{r}_{i}\partial_{j}\zeta-\hat{r}_{j}\partial_{i}\zeta\right]e^{-i\omega\zeta\hat{n}\cdot\hat{r}},\label{tij1-1}\\
	t_{ij}^{(2)} & =\left[\partial_{i}\zeta\partial_{j}\zeta+\hat{r}_{i}\hat{r}_{j}\left(\frac{\zeta^{2}}{R^{2}}-\frac{(\nabla\zeta)^{2}}{2}\right)-2\frac{\zeta}{R}\left(\hat{r}_{i}\partial_{j}\zeta+\hat{r}_{j}\partial_{i}\zeta\right)\right]e^{-i\omega\zeta\hat{n}\cdot\hat{r}},\label{tij2-1}
\end{align}
and so on. We did not expand the exponential $e^{-i\omega\zeta\hat{n}\cdot\hat{r}}$,
since $\omega\zeta$ may be large depending on $ \omega $. We shall 
consider for simplicity the case in which the length
scale $k_{*}^{-1}$ of the deformations is well separated from the typical
bubble radius $ R_* $. Therefore, we have $ k_* R_*\gg 1 $ and $\partial_i\zeta\sim k_{*}\zeta\gg\zeta/R$,
which simplifies the expressions, 
\begin{align}
	t_{ij}^{(0)} & =\hat{r}_{i}\hat{r}_{j}e^{-i\omega\zeta\hat{n}\cdot\hat{r}},   \label{tij0}   \\
	t_{ij}^{(1)} & =-\left(\hat{r}_{i}\partial_{j}\zeta+\hat{r}_{j}\partial_{i}\zeta\right)e^{-i\omega\zeta\hat{n}\cdot\hat{r}},\label{tij1}\\
	t_{ij}^{(2)} & =\left[\partial_{i}\zeta\partial_{j}\zeta-\hat{r}_{i}\hat{r}_{j}(\nabla\zeta)^{2}/2\right]e^{-i\omega\zeta\hat{n}\cdot\hat{r}}.
	\label{tij2}
\end{align}
We shall assume $ k_*\zeta\lesssim 1 $ and expand the product $t_{ij}t_{kl}^{\prime*}$ to quadratic order in $\partial_i\zeta$, so we have
\begin{equation}
	\langle t_{ij}t_{kl}^{\prime*}\rangle = 
	\langle t_{ij}^{(0)}t_{kl}^{\prime(0)} \rangle 
	+\langle t_{ij}^{(0)}t_{kl}^{\prime*(1)}\rangle +\langle t_{ij}^{(1)}t_{kl}^{\prime*(0)}\rangle 
	+\langle  t_{ij}^{(0)} t_{kl}^{\prime(2)}\rangle^{*} + \langle t_{ij}^{(2)} t_{kl}^{\prime(0)}\rangle 
	+ \langle t_{ij}^{(1)}t_{kl}^{\prime*(1)}\rangle. \label{tt2}
\end{equation}
In computing this quantity, we will have to deal with the average
\begin{equation}
	\langle e^{i\omega\hat{n}\cdot(\zeta'\hat{r}'-\zeta\hat{r})}\rangle =
	\exp\left\{ -\textstyle{\frac{1}{2}}\omega^{2}[(\hat{n}\cdot\hat{r})^{2}\langle \zeta^{2}\rangle 
	+(\hat{n}\cdot\hat{r}')^{2}\langle \zeta^{\prime2}\rangle -2(\hat{n}\cdot\hat{r})(\hat{n}\cdot\hat{r}')\langle \zeta\zeta'\rangle ]\right\}.
	\label{expm}
\end{equation} 
as well as
\begin{align}
	\langle \partial_{i}\zeta e^{i\omega\hat{n}\cdot(\zeta'\hat{r}'-\zeta\hat{r})}\rangle 
	&= 
	i\omega(\hat{n}\cdot\hat{r}') \langle \partial_{i}\zeta\zeta'\rangle 
	\langle e^{i\omega\hat{n}\cdot(\zeta'\hat{r}'-\zeta\hat{r})}\rangle ,  
	\label{dizm}
\\
	\langle \partial_{i}\zeta\partial_{j}\zeta e^{i\omega\hat{n}\cdot(\zeta'\hat{r}'-\zeta\hat{r})}\rangle 
	&=
	[\langle \partial_{i}\zeta\partial_{j}\zeta\rangle -\omega^2(\hat{n}\cdot\hat{r}')^{2}\langle \zeta'\partial_{i}\zeta\rangle 
	\langle \zeta'\partial_{j}\zeta\rangle ] 
	\langle e^{i\omega\hat{n}\cdot(\zeta'\hat{r}'-\zeta\hat{r})}\rangle ,
	\label{dizdjzm}
\\
	\langle \partial_{i}\zeta\partial_{k}\zeta'e^{i\omega\hat{n}\cdot(\zeta'\hat{r}'-\zeta\hat{r})}\rangle 
	&= 
	[\langle \partial_{i}\zeta\partial_{k}\zeta'\rangle -\omega^2(\hat{n}\cdot\hat{r})(\hat{n}\cdot\hat{r}')
	\langle \zeta'\partial_{i}\zeta\rangle \langle \zeta\partial_{k}\zeta'\rangle] 
	\langle e^{i\omega\hat{n}\cdot(\zeta'\hat{r}'-\zeta\hat{r})}\rangle 
	\label{dizdkzpm}
\end{align}
(and similar expressions interchanging $ \hat r\leftrightarrow\hat r ' $ and $ \zeta\leftrightarrow\zeta ' $).
We derive these expressions in appendix \ref{apave} assuming Gaussian fluctuations.
For the two-bubble contribution we have $\left\langle \zeta\zeta'\right\rangle =0$, and several terms in these expressions vanish.

The zeroth order term in the expansion of $  t_{ij}t_{kl}^{\prime*}$ gives
\begin{equation}
	\Lambda_{ij,kl}\langle t_{ij}^{(0)}t_{kl}^{\prime*(0)}\rangle 
	=(\Lambda_{ij,kl}\hat{r}_{i}\hat{r}_{j}\hat{r}'_{k}\hat{r}'_{l})
	\langle e^{i\omega\hat{n}\cdot(\zeta'\hat{r}'-\zeta\hat{r})}\rangle .
	\label{ordencero}
\end{equation}
The first factor, $\Lambda_{ij,kl}\hat{r}_{i}\hat{r}_{j}\hat{r}'_{k}\hat{r}'_{l}$, is given by Eq.~(\ref{Ltt0}).
This factor alone would give the result of 
the bubble collision mechanism. However, it is modified by the average $ \langle e^{i\omega\hat{n}\cdot(\zeta'\hat{r}'-\zeta\hat{r})}\rangle $. 
Since we consider small deformations, such that $ \zeta\ll R_* $, for frequencies $ \omega\sim R_*^{-1} $ 
Eq.~(\ref{expm}) gives $ \langle e^{i\omega\hat{n}\cdot(\zeta'\hat{r}'-\zeta\hat{r})}\rangle\sim 1 $.
Therefore, Eq.~(\ref{ordencero}) will reproduce the peak of the envelope approximation.
If both the size and the amplitude of the  relevant deformations are characterized by the length scale $ k_*^{-1} $, 
we have  $ \langle\zeta^2 \rangle\sim k_*^{-2}$, and the argument of the exponential (\ref{expm}) will become of order 1 at the higher scale $ \omega\sim k_* $. 

Using the property $\Lambda_{ij,kl}=\Lambda_{ji,lk}$ and Eq.~(\ref{dizm}),
the linear order terms in the expansion of $  t_{ij}t_{kl}^{\prime*}$ give 
\begin{multline}
	\Lambda_{ij,kl}\left(\langle t_{ij}^{(0)}t_{kl}^{\prime*(1)}\rangle +\langle t_{ij}^{(1)}t_{kl}^{\prime*(0)}\rangle \right)=
	\\  2i\omega\Lambda_{ij,kl}
	\left[(\hat{n}\cdot\hat{r})\hat{r}_{i}\hat{r}_{j}\hat{r}_{k}'\langle\zeta\partial_{l}\zeta'\rangle-(\hat{n}\cdot\hat{r}')\hat{r}_{i}\hat{r}_{k}'\hat{r}_{l}'\langle\partial_{j}\zeta\zeta'\rangle\right]\langle e^{i\omega\hat{n}\cdot(\zeta'\hat{r}'-\zeta\hat{r})}\rangle 
	\label{linearorder}
\end{multline}
For the two-bubble contribution these terms vanish, since we have
$\left\langle \zeta\partial_{l}\zeta'\right\rangle =\left\langle \zeta\right\rangle \left\langle \partial_{l}\zeta'\right\rangle =0$ and the same for $ \langle\partial_{j}\zeta\zeta'\rangle $.
In the single-bubble case and for $ k_*\zeta\lesssim 1 $, these terms are small for $ \omega\sim R_*^{-1} $, but are in principle of order 1 for $ \omega\sim k_* $.
However, one expects that short angular separations $ \sim k_*^{-1}/R_* $ are relevant, and 
these averages are small for $\hat{r}\simeq\hat{r}'$, as shown in Eq.~(\ref{zetadzetap}).

Let us now consider one of the three second-order terms contributing to $ \left\langle t_{ij}t_{kl}^{\prime*}\right\rangle  $, 
\begin{multline}
	\Lambda_{ij,kl}\langle t_{ij}^{(2)}t_{kl}^{\prime(0)*}\rangle =
	\Lambda_{ij,kl}\,\hat{r}_{k}'\hat{r}_{l}'\left[\langle \partial_{i}\zeta\partial_{j}\zeta\rangle -\omega^{2}(\hat{n}\cdot\hat{r}')^{2}\langle \zeta'\partial_{i}\zeta\rangle \langle \zeta'\partial_{j}\zeta\rangle \right]
	\langle e^{i\omega\hat{n}\cdot(\zeta'\hat{r}'-\zeta\hat{r})}\rangle 
	\\
	-\Lambda_{ij,kl}\,\hat{r}_{i}\hat{r}_{j}\hat{r}_{k}'\hat{r}_{l}'\,\textstyle{\frac{1}{2}}\left[\langle (\nabla\zeta)^{2}\rangle -\omega^{2}(\hat{n}\cdot\hat{r}')^{2}\langle \zeta'\nabla\zeta\rangle ^{2}\right]
	\langle e^{i\omega\hat{n}\cdot(\zeta'\hat{r}'-\zeta\hat{r})}\rangle .
	\label{secondorder20}
\end{multline}
where we have used Eq.~(\ref{dizdjzm}). The term $\langle t_{ij}^{(0)}t_{kl}^{\prime(2)*}\rangle $
is similar, with $\hat{r}\leftrightarrow\hat{r}'$ and $\zeta\leftrightarrow\zeta'$.
We have  not neglected the terms of order $\zeta^4$, since
the accompanying factor $\omega^{2}$ may be large.
Nevertheless, as discussed above, for the two-bubble contribution  the averages $\left\langle \zeta'\partial_{i}\zeta\right\rangle $ vanish, and
in the single-bubble case they become small for deformations in a small length scale.
Finally, for the term $ \langle t_{ij}^{(1)}t_{kl}^{\prime(1)*}\rangle  $ we have, using Eq.~(\ref{dizdkzpm}) and the property $ \Lambda_{ij,kl}=\Lambda_{ji,lk} $,
\begin{multline}
	\Lambda_{ij,kl}\langle t_{ij}^{(1)}t_{kl}^{\prime(1)*}\rangle =
	2\Lambda_{ij,kl}\left[\hat{r}_{j}\hat{r}_{k}'\langle \partial_{i}\zeta\partial_{l}\zeta'\rangle +\hat{r}_{j}\hat{r}_{l}'\langle \partial_{i}\zeta\partial_{k}\zeta'\rangle \right]
	\langle e^{i\omega\hat{n}\cdot(\zeta'\hat{r}'-\zeta\hat{r})}\rangle
	\\
	+2\omega^{2}(\hat{n}\cdot\hat{r})(\hat{n}\cdot\hat{r}')\Lambda_{ij,kl}\left[\hat{r}_{j}\hat{r}_{k}'\langle \zeta'\partial_{i}\zeta\rangle \langle \zeta\partial_{l}\zeta'\rangle 
	+\hat{r}_{j}\hat{r}_{l}'\langle \zeta'\partial_{i}\zeta\rangle \langle \zeta\partial_{k}\zeta'\rangle \right]\langle e^{i\omega\hat{n}\cdot(\zeta'\hat{r}'-\zeta\hat{r})}\rangle .
	\label{secondorder11}
\end{multline}
This whole expression vanishes for the two-bubble case. 
For the single-bubble contribution, the terms which are of order $\omega^2\zeta^4$ (the whole last line), 
will be very small for a small length scale, 
due to small averages $\left\langle \zeta'\partial_{i}\zeta\right\rangle $.

\subsection{Contributions to $ \Pi $}

To avoid long expressions, from now on we shall neglect the terms in (\ref{linearorder})-(\ref{secondorder11}) which are small when
the characteristic length $ k_*^{-1} $ is such that $ k_* R_*\gg 1 $. Including
them in a calculation (if necessary) is straightforward.
Hence, we have
\begin{align}
	\Lambda_{ij,kl}\left\langle t_{ij}t_{kl}^{\prime*}\right\rangle \simeq\, 
	& \Lambda_{ij,kl} \, \left\{ \hat{r}_{i}\hat{r}_{j}\hat{r}'_{k}\hat{r}'_{l}\right.\nonumber \\
	& +\hat{r}_{k}'\hat{r}_{l}'\langle \partial_{i}\zeta\partial_{j}\zeta\rangle +\hat{r}_{i}\hat{r}_{j}\langle \partial_{k}\zeta'\partial_{l}\zeta'\rangle 
	-\hat{r}_{i}\hat{r}_{j}\hat{r}_{k}'\hat{r}_{l}' \, \textstyle{\frac{1}{2}}\left[\langle (\nabla\zeta)^{2}\rangle + \langle (\nabla\zeta')^{2}\rangle \right] \nonumber \\
	& \left.+2\hat{r}_{j}\hat{r}_{k}'\langle \partial_{i}\zeta\partial_{l}\zeta'\rangle +2\hat{r}_{j}\hat{r}_{l}'\langle \partial_{i}\zeta\partial_{k}\zeta'\rangle \right\} 
	\,\langle e^{i\omega\hat{n}\cdot(\zeta'\hat{r}'-\zeta\hat{r})}\rangle.
	\label{contribuciones}
\end{align}
Inside the brackets, the first term is the contribution from $t_{ij}^{(0)}t_{kl}^{\prime(0)*}$,
the terms in the second line are those from $t_{ij}^{(0)}t_{kl}^{\prime(2)*}$
and $t_{ij}^{(2)}t_{kl}^{\prime(0)*}$, and the last line contains the contribution
from $t_{ij}^{(1)}t_{kl}^{\prime(1)*}$. We shall denote the corresponding
contributions to $\Pi$ by $\Pi^{(0,0)}$, $\Pi^{(0,2)}$, $\Pi^{(2,0)}$,
and $\Pi^{(1,1)}$, respectively.
Each of these terms separates into a single-bubble contribution corresponding to Eq.~(\ref{Pissepar}) and a two-bubble contribution 
corresponding to Eq.~(\ref{Pidsepar}),
except for $\Pi^{(1,1)}$, for which the two-bubble contribution vanishes.
Like in the spherical case, the quantity $ \sigma $ varies during the phase transition, 
and is related to the energy released by each bubble. 
We shall assume again that it is given by Eq.~(\ref{sigmarhovac}). 
Thus,  all the terms $\Pi^{(0,0)}$, $\Pi^{(0,2)}$, $\Pi^{(2,0)}$,
$\Pi^{(1,1)}$ have a common factor 
\begin{equation}
	\sigma\sigma'=(\kappa\rho_{\mathrm{vac}})^{2}RR'/9.\label{sigmaR}
\end{equation}

According to Eq.~(\ref{contribuciones}), for the zeroth-order term 
Eqs.~(\ref{Pissepar})-(\ref{Pidsepar}) coincide with Eqs.~(\ref{Piscol})-(\ref{Pidcol}), except for the 
factor
\begin{equation}
	\langle e^{i\omega\hat{n}\cdot(\zeta'\hat{r}'-\zeta\hat{r})}\rangle =
	\exp\left\{ -\textstyle{\frac{1}{2}}\omega^{2}[\alpha_0(\hat{n}\cdot\hat{r})^{2}
	+\alpha_0'(\hat{n}\cdot\hat{r}')^{2} - 2\alpha(\hat{n}\cdot\hat{r})(\hat{n}\cdot\hat{r}') ]\right\},
	\label{expalfa}
\end{equation} 
where we use the notation $ \alpha_0\equiv \alpha_{0}(t) $, $ \alpha_0'=\alpha_0(t') $, $ \alpha=\alpha(\theta,t,t') $. We have
\begin{multline}
	\Pi^{(0,0)}  = 
	\frac{\kappa^{2}\rho_{\mathrm{vac}}^{2}}{9}  \Big[	\int_{-\infty}^{t}dt_{N}\Gamma(t_{N})R^{3}R^{\prime 3}
	\int d\hat{r}\int d\hat{r}'\,e^{-I_{\mathrm{tot}}}e^{i\omega\hat{n}\cdot\mathbf{s}}A\, G^{(s)}   
	\\
	+\int_{-\infty}^{t}dt_{N}\Gamma(t_{N})\int_{-\infty}^{t'}dt_{N}^{\prime}\Gamma(t_{N}^{\prime})R^{3}R^{\prime 3}
	\int d^{3}s e^{-I_{\mathrm{tot}}} e^{i\omega\hat{n}\cdot\mathbf{s}}  \int d\hat{r}\int d\hat{r}'
	A\, G^{(d)} \Big].
		\label{Pi00}
\end{multline}
where $ A=\Lambda_{ij,kl}\hat{r}_{i}\hat{r}_{j}\hat{r}'_{k}\hat{r}'_{l} $ and
$ G^{(s)} $ and $ G^{(d)} $ are given by Eq.~(\ref{expalfa}), only that for the latter we have $ \alpha=0 $ [also, the angular integrals in the second line are constrained by the conditions (\ref{cM})-(\ref{cMp})].
As anticipated, Eq.~(\ref{Pi00}) is very similar to the bubble collision case  (\ref{Piscol})-(\ref{Pidcol}). 
The main difference is the presence of the Gaussians in the integrand\footnote{Notice that this extra dependence hinders any attempt to integrate out the vector $ \hat n $ like we did for the bubble collision mechanism.}.
For the contribution $ \Pi^{(0,2)} + \Pi^{(2,0)} $, we notice that the second line of Eq.~(\ref{contribuciones}), 
using the last of Eqs.~(\ref{correlssamepoint}) and taking into account that $ \Lambda_{ij,kl}\delta_{ij}=0 $, gives 
\begin{equation}
-2[ \beta_{0}(t) + \beta_{0}(t') ] \Lambda_{ij,kl} \, \hat{r}_{i}\hat{r}_{j}\hat{r}_{k}'\hat{r}_{l}' ,
\label{correc2beta0}
\end{equation}
which is of the same form of the previous contribution, and we obtain
\begin{equation}
	\Pi^{(0,0)}+\Pi^{(2,0)}+\Pi^{(0,2)}=\left[1-2\beta_{0}(t) -2\beta_{0}(t') \right] \Pi^{(0,0)},
	\label{Pi002002}
\end{equation}

On the other hand, for the contribution $\Pi^{(1,1)}$, we use Eq.~(\ref{dzetadzetap}) in the last line of Eq.~(\ref{contribuciones})
(in this case the two-bubble contribution vanishes). Using the property $ \Lambda_{ij,kl}=\Lambda_{ji,lk} $, we obtain 
\begin{multline}
	\Lambda_{ij,kl}\left(2\hat{r}_{j}\hat{r}_{l}'\langle\partial_{i}\zeta\partial_{k}\zeta'\rangle+2\hat{r}_{j}\hat{r}_{k}'\langle\partial_{i}\zeta\partial_{l}\zeta'\rangle\right)=\\
	2\Lambda_{ij,kl}\frac{\hat{r}_{j}\left(\hat{r}_{l}'\delta_{ik}+\hat{r}_{k}'\delta_{il}\right)-2\hat{r}_{i}\left(\hat{r}_{j}\hat{r}_{k}+\hat{r}_{j}'\hat{r}_{k}'\right)\hat{r}_{l}'+2c\hat{r}_{i}\hat{r}_{j}\hat{r}_{k}'\hat{r}_{l}'}{RR'}\frac{\partial\alpha}{\partial c}  
	\\
	+2\Lambda_{ij,kl}\frac{\hat{r}_{i}'\hat{r}_{j}\left(\hat{r}_{l}'\hat{r}_{k}+\hat{r}_{k}'\hat{r}_{l}\right)+2c^{2}\hat{r}_{i}\hat{r}_{j}\hat{r}_{k}'\hat{r}_{l}'-2c\hat{r}_{i}\left(\hat{r}_{j}\hat{r}_{k}+\hat{r}_{j}'\hat{r}_{k}'\right)\hat{r}_{l}'}{RR'}\frac{\partial^{2}\alpha}{\partial c^{2}}
	\label{Ltt11}
\end{multline}
In this case, besides the  expression $A= \Lambda_{ij,kl}\hat{r}_{i}\hat{r}_{j}\hat{r}_{k}'\hat{r}_{l}' $, which is given by Eq.~(\ref{Ltt0}), we have also
\begin{align}
	A_1 &=\Lambda_{ij,kl}\left(\hat{r}_{i}'\hat{r}_{j}\hat{r}_{k}\hat{r}_{l}'+\hat{r}_{i}'\hat{r}_{j}\hat{r}_{k}'\hat{r}_{l}\right) = \left[1-(\hat{n}\cdot\hat{r})^{2}\right]\left[1-(\hat{n}\cdot\hat{r}')^{2}\right],
	\label{termino1}
\\
	A_2 &=\Lambda_{ij,il}\hat{r}_{j}\hat{r}_{l}' + \Lambda_{ij,ki}\hat{r}_{j}\hat{r}_{k}' = 2\left[\hat{r}\cdot\hat{r}'-(\hat{n}\cdot\hat{r})(\hat{n}\cdot\hat{r}')\right],
\\
	A_3 &=\Lambda_{ij,kl}\hat{r}_{j}\hat{r}_{l}'\left(\hat{r}_{i}\hat{r}_{k}+\hat{r}_{i}'\hat{r}_{k}'\right)=
	\textstyle{\frac{1}{2}}\left[(r\cdot r')-(\hat{n}\cdot r)(\hat{n}\cdot r')\right]\left[2-(\hat{n}\cdot r)^{2}-(\hat{n}\cdot r')^{2}\right].
	\label{termino3}
\end{align}
Thus, the contribution to Eq.~(\ref{Pissepar}) is of the form
\begin{equation}
	\Pi^{(1,1)} = \frac{\kappa^{2}\rho_{\mathrm{vac}}^{2}}{9}  \int_{-\infty}^{t}dt_{N}\Gamma(t_{N})R^{3}R^{\prime 3} 
	\int d\hat{r}\int d\hat{r}'\,e^{-I_{\mathrm{tot}}}e^{i\omega\hat{n}\cdot\mathbf{s}}G^{(s)} \frac{B\partial_c \alpha+C \partial_c^{2}\alpha}{RR'},
	\label{Pi11}
\end{equation} 
with $ B=2\left(A_{2}-2A_{3}+2cA\right) $ and $ C=2\left(A_{1}+2c^{2}A-2cA_{3}\right) $.

\section{A single deformation scale}
\label{toymodel}

The GW spectrum resulting from Eqs.~(\ref{Pi00}), (\ref{Pi002002}), (\ref{Pi11}) will depend 
on the dynamics of the deformations. 
For instance, in the case of hydrodynamic instabilities, 
the coupled equations for the wall deformations and the fluid perturbations must be considered. These quantities
can be expanded in modes which diagonalize the equations.
This has been done for a planar wall, in which case 
there is a range of unstable wavenumbers, and
it is to be expected that the situation is similar in the spherical case.
The most unstable wavelength
is in general much shorter than the typical bubble radius by the end of the phase transition (see, e.g., \cite{mm14}). 
The evolution of these perturbations beyond the linear order has not been studied, at least in this context\footnote{See e.g.\ \cite{BychkovLiberman} for similar instabilities in flames.}. 
One possibility is that the most unstable mode dominates, and we shall consider for simplicity a single length scale.
If the instability persists in the nonlinear regime, this may lead to dendritic growth \cite{fa90}. 
We will be more conservative and  assume that the deformation grows up to an amplitude of the order of the deformation size $ k_*^{-1} $.

\subsection{The model}

\begin{figure}[bt]
	\centering
	\includegraphics[width=\textwidth]{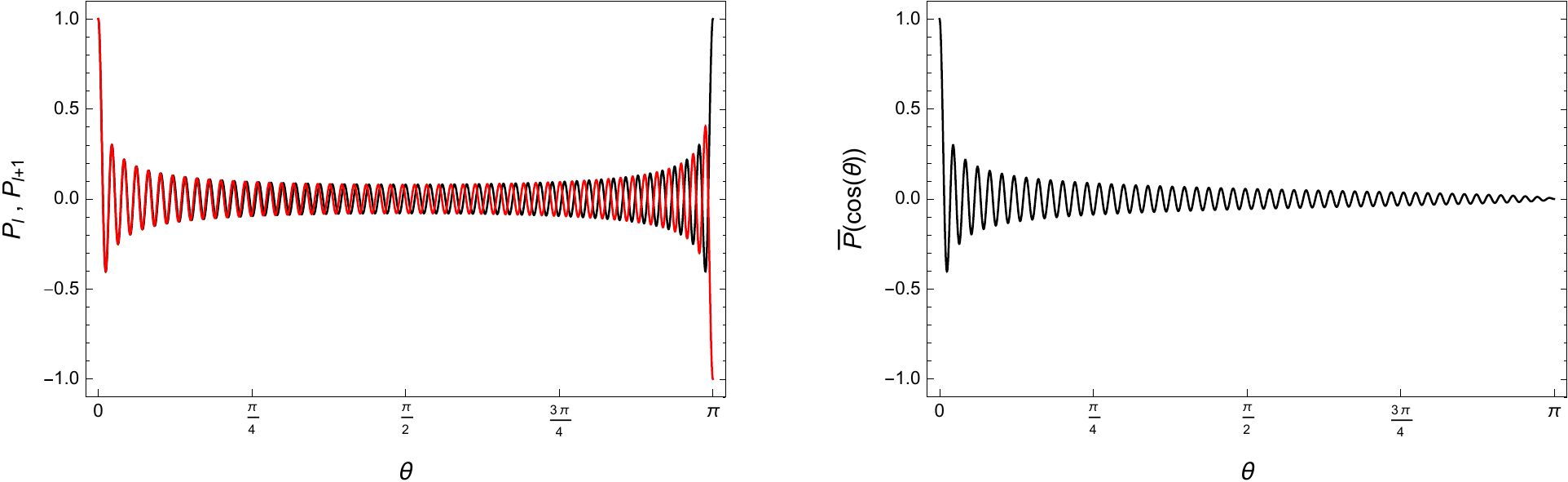}
	\caption{Two adjacent Legendre polynomials $ P_l$ (black) and $P_{l+1} $ (red) with $ l=100 $ and the function $ \bar P(c) $ for $ l_* =100$.}
	\label{figpl}
\end{figure}

We could model such deformations by considering a single component in the expansion (\ref{alfa}), i.e., setting
$ |D|_{l}^{2}(t,t')\propto \delta_{ll_{*}} $, 
where $ l_* $ is related to the deformation scale by $ k_{*}^{-1}\sim R_*/l_{*} $.
Assuming  an average displacement $\sqrt{\langle\zeta^{2}\rangle}\sim k_*^{-1} $, Eq.~(\ref{alfa}) gives 
%[using $ P_{l}(1)=1 $] 
\begin{equation}
	\frac{2l+1}{4\pi}|D|_{l}^{2}(t,t)=\left(\frac{a}{l_{*}}\right)^{2}R^{2}\,\delta_{ll_{*}}.
\end{equation}
where $ a $ is a constant of order 1.
For the unequal-time correlation function, we may assume for simplicity a maximal time correlation, 
$ |D|_{l}(t,t')=\sqrt{|D|_{l}^{2}(t,t)|D|_{l}^{2}(t',t')} $,
which gives
\begin{equation}
	\frac{2l+1}{4\pi}|D|_{l}^{2}(t,t')=\left(\frac{a}{l_{*}}\right)^{2}RR'\delta_{ll_{*}}.
	\label{fcorr}
\end{equation}
Thus, Eq.~(\ref{alfa}) becomes
$\alpha(c,t,t')=({a}/{l_{*}})^{2}RR'P_{l_{*}}(c)$.
The function $ P_{l_*} (c)$  has a maximum at $ c=1 $ (i.e., at $ \theta=0 $), with $ P_{l_*}(1)=1 $. 
For $ 0<\theta <\pi $, 
this function oscillates with a smaller amplitude, but it has another extremum at $ \theta=\pi $, with $ P_{l_*}(-1)=(-1)^{l_*} $.
This second peak of the correlation function for opposite points on the bubble wall is rather artificial. If we considered a displacement spectrum with many components $ l $, the correlation function would actually  vanish for $ c=-1 $ due to the superposition of alternating signs. 
We may reproduce this effect without departing significantly from the simple model (\ref{fcorr}) by considering two adjacent components,
\begin{equation}
	\alpha(c,t,t')=\frac{a^2}{l_{*}^2}RR'\frac{P_{l_{*}}(c)+P_{l_{*}+1}(c)}{2}\equiv \frac{a^2}{l_{*}^2}RR'\bar P(c).
	\label{alfaPl}
\end{equation}
Figure \ref{figpl} shows the form of the function $ \bar P(c) $ for $ l_*=100 $.

We shall consider the case $ l_*\gg 1 $, so that we have $ k_*^{-1}\ll R_* $.  
Thus, Eq.~(\ref{alfaPl}) gives $ \langle\zeta^{2}\rangle=\alpha(1,t,t) \sim (R_*/l_*)^2\ll 1$, which justifies the expansion of Eq.~(\ref{tij})
in powers of $ \zeta/R $ and the approximations 
used in Eqs.~(\ref{tij0})-(\ref{tij2}).
On the other hand, the derivatives of $ \alpha $ are much larger. In particular, 
taking into account that $ P_l'(1)=l(l+1)/2 $, we have $\beta_0=R^{-2}\partial_c\alpha(1,t,t)\simeq a^2/2$.  
Since $\langle(\nabla\zeta)^{2}\rangle=2\beta_{0}(t)\simeq a^2$, the constant $ a $ must be smaller than 1 so that the expansion of Eq.~(\ref{tij}) in powers of $ (\nabla\zeta)^{2} $ is valid.
A well known asymptotic approximation for the Legendre polynomials for large $ l_* $, namely, 
$ P_{l_{*}}(\cos\theta)= \sqrt{\frac{2}{\pi l_{*}\sin\theta}}\cos\left[\left(l_{*}+\frac{1}{2}\right)\theta-\frac{\pi}{4}\right]$, gives
\begin{equation}
	\bar{P}(\cos\theta)\simeq \sqrt{\frac{2}{\pi l_{*}\sin\theta}}\cos\frac{\theta}{2}\cos\left(l_{*}\theta-\frac{\pi}{4}\right),
	\label{aproxlgrande}
\end{equation} 
which is a very good approximation for $ \theta\geq l_*^{-1} $. 
For $ \theta< l_*{-1} $, we may use the Taylor expansion $ 1-\frac{1}{4}l_{*}^{2}\theta^{2}+\frac{1}{64}l_{*}^{4}\theta^{4}$
(where we have also made the approximation of large $ l_* $).

We will also consider for simplicity  a constant velocity $ v $ and  a simultaneous nucleation, i.e., $ \Gamma(t)=n_* \delta(t-t_*) $, 
where $ n_* $ is the number density of bubbles.
We use this parameter to  define the characteristic bubble size by $ R_* = n_*^{-1/3} $, and we define a dimensionless  spectrum\footnote{A 
	very similar function was defined in Refs.~\cite{hk08,jt17} for an exponential nucleation rate $ \Gamma=\Gamma_* e^{\beta(t-t_*)} $, with $ R_*^{-2} $ replaced by $ \beta^2 $.} 
\begin{equation}
	\Delta(R_*\omega)= \frac{3R_*^{-2}}{8\pi G\kappa^{2}\rho_{\mathrm{vac}}^{2} }\frac{d\rho_{GW}}{d\ln\omega} (\omega).
	\label{defDelta}
\end{equation}
Inserting Eqs.~(\ref{Pi00}) and (\ref{Pi11}) in Eq.~(\ref{rhogw}), we obtain the contributions to $ \Delta $. 
With the change of variables $ t,t'\to t-t_*,t'-t_* $	 we have
%\begin{equation}
%	\Delta =\frac{3\omega^{3} R_*^{-2}}{2\pi^2 \kappa^{2}\rho_{\mathrm{vac}}^{2} }\int_{-\infty}^{\infty}dt\int_{t}^{\infty}dt'\cos[\omega(t-t')]\,\Pi(t,t',\omega),
%\end{equation}
\begin{multline}
	\Delta^{(0,0)} =	 \frac{\omega^{3}}{3\pi R_*^{5}}
	\int_{0}^{\infty}dt\int_{t}^{\infty}dt'\cos[\omega (t'-t)] (RR')^3
	\\ \times 
	 \left[\int d\hat r \int d\hat r'
		e^{-I_{\mathrm{tot}}}e^{i\omega\hat{n}\cdot\mathbf{s}}A\,G^{(s)}   \right.
	+ \left. R_*^{-3}	\int d^{3}s\,	 e^{-I_{\mathrm{tot}}} e^{i\omega\hat{n}\cdot\mathbf{s}}
	\int d\hat r \int d\hat r'
	 A\,G^{(d)} \right],
	 \label{Delta00}
\end{multline}
where $ G^{(s,d)} $ denote the Gaussian (\ref{expalfa})  for the single-bubble or two-bubble case,
\begin{align}
	G^{(s)} &=	\exp\left\{ - \frac{\omega^{2}a^2}{2l_*^2} \left[(\hat{n}\cdot\hat{r})^{2}R^2 + (\hat{n}\cdot\hat{r}')^{2} R^{\prime2}
	- 2(\hat{n}\cdot\hat{r})(\hat{n}\cdot\hat{r}') RR' \bar P(c) \right]\right\},
	\label{gaussianas}
\\
	G^{(d)} &= \exp\left\{ - \frac{\omega^{2}a^2}{2l_*^2} \left[(\hat{n}\cdot\hat{r})^{2}R^2 + (\hat{n}\cdot\hat{r}')^{2} R^{\prime2} \right]\right\}.
	\label{gaussianad}
\end{align}	
In terms of $ R_\pm=vt_\pm =v(t'\pm t)$ we have
\begin{equation}
I_{\mathrm{tot}}=\frac{\pi}{12R_{*}^{3}}\left[2R_{+}^{3}+3sR_{+}^{2}-s^{3}+3s^{-1}\left(R_{+}+s\right)^{2}R_{-}^{2}\right],
\end{equation} 
with $ s^2 =R^2+R^{\prime 2}-2RR'c $ for the single-bubble term, while $ \mathbf{s} $ is an independent variable for
the two-bubble term. In the latter, the integrals are restrained by the conditions
\begin{equation}
	\hat{r}\cdot\mathbf{s}  \geq-\frac{s^{2}-R_+R_-}{R_+-R_-},\quad
	\hat{r}'\cdot\mathbf{s}  \leq\frac{s^{2}+R_+R_-}{R_++R_-}.
\end{equation}
On the other hand, for this model  Eq.~(\ref{Pi002002}) gives 
\begin{equation}
	\Delta^{(0,0)}+\Delta^{(2,0)}+\Delta^{(0,2)}=\left(1-2a^2\right) \Delta^{(0,0)}.
\end{equation}

If we integrate first $ \hat r' $, putting $ \hat r $ in the $ z $ axis and $ \hat n $ in the $ xz $ plane, we have 
\begin{equation}
	\hat r'\cdot \hat r = \cos\theta, \quad
	 \hat n\cdot\hat r=\cos \chi , \quad
	\hat n\cdot\hat r' = \cos\chi  \cos\theta+\sin\chi\cos\phi \sin\theta , 
	\label{cnp}
\end{equation} 
where $ \theta $ is the angle between $ \hat r' $ and $ \hat r $, 
$ \phi $ is the azimuthal angle of $ \hat r ' $, and
$ \chi $ is the angle between $ \hat r $ and $ \hat n $ (see Fig.~\ref{figangles}). Thus,
the angular integrations become $ \int d\hat r \int d\hat r'= 2\pi\int_{-1}^{+1} dc_\chi \int_{-1}^{+1}dc \int_0^{2\pi} d\phi$. 
Using the abbreviated notation $ c_\chi = \cos\chi$, $ s_\chi=\sin\chi $, $ c_\phi = \cos\phi $ (as well as $ c=\cos\theta $), we have
\begin{equation}
	A=\frac{1}{2}s_\chi^2\left[2(cs_{\chi}-c_{\chi}c_{\phi}\sqrt{1-c^2})^{2}+(c_\chi c+s_\chi c_\phi \sqrt{1-c^2})^{2}-1 \right].
\end{equation}
Finally,  Eq.~(\ref{Pi11}) gives (changing variables $ t,t'\to R,R' $)
\begin{multline}
	\Delta^{(1,1)} = \frac{a^2\omega^{3} R_*^{-2}}{3\pi l_*^2 v^2} 
	\int_{0}^{\infty}dR\int_{R}^{\infty}dR'\cos(\omega R_-/v)	 R^{3}R^{\prime 3} 
	\int_{-1}^{+1} dc_\chi \int_0^{2\pi} d\phi
	\\  \times  
	\int_{-1}^{+1}dc  \,
	e^{-I_{\mathrm{tot}}} \, e^{i\omega\hat{n}\cdot\mathbf{s}}G^{(s)} 
	\left(B\partial_c \bar P+ C \partial_c^{2}\bar P\right),
	\label{Pi11P}
\end{multline} 
with 
\begin{align}
	B &	=2\left[2(1-s_{\chi}^{2}c_{\phi}^{2})s_{\chi}^{2}c^{3}+(2c_{\phi}^{2}-1)s_{\chi}^{4}c
	+c_{\chi}s_{\chi}c_{\phi}(s_{\chi}^{2}c_{\phi}^{2}c^{2}-c_{\chi}^{2}c^{2}-c_{\chi}^{2}-s_{\chi}^{2}c_{\phi}^{2})\sqrt{1-c^{2}}\right], \nonumber
	\\
	C &	=2\left[2(s_{\chi}^{2}c_{\phi}^{2}-1)s_{\chi}^{2}c^{2}-(s_{\chi}^{2}c_{\phi}^{2}-c_{\chi}^{2})c_{\chi}s_{\chi}c_{\phi}c\sqrt{1-c^{2}}
	+s_{\chi}^{2}(1-s_{\chi}^{2}c_{\phi}^{2})\right]\left(1-c^{2}\right),
	\label{AB}
\end{align}
and $\hat n\cdot\mathbf{s}= \left(R'\cos\theta-R\right) c_\chi +R' s_\chi\,c_\phi \,\sin\theta $.

In order to deal with the function $ \bar P(\cos\theta) $ in the exponent of Eq.~(\ref{gaussianas}), we notice that, for large $ l_* $, this function takes order-1 values only in a very small interval of length $ l_*^{-1} $ close to $ \theta=0 $, while for most of the interval $ 0<\theta<\pi $ it oscillates with an amplitude of order $ l_*^{-1/2} $. Therefore, we can write
\begin{equation}
	G^{(s)} \simeq G^{(d)}	\left[1+ (\omega a/l_*)^2 c_\chi(c_\chi  \cos\theta + s_\chi\cos\phi \sin\theta  ) RR'\bar P(c)\right],
	\label{gaussap0}
\end{equation} 
This approximation is valid for values of $ \omega $ up to $ \omega\sim l_*/R_* $, while for $ \omega \gg l_*/R_* $ the quantity $ \Delta $ will be negligible anyway.
According to the approximation (\ref{aproxlgrande}), the function $ \bar P $ in the second term of Eq.~(\ref{gaussap0}) combines with $\partial_c \bar P $ 
and $\partial_c^2 \bar P $ in Eq.~(\ref{Pi11P}) to form an oscillating function with frequency $ 2l_* $.
Therefore, the calculation of this contribution is similar to that of the first term.
We shall ignore this $ \mathcal{O}(l_* ^{-1/2}) $ correction for simplicity, so from now on we assume  
\begin{equation}
	G^{(s)} \simeq G^{(d)} =  \exp \left\{-\frac{1}{2}\left(\frac{\omega a}{l_*}\right)^2 \left[c_\chi^{2}R^2 + (c_\chi  \cos\theta + s_\chi\cos\phi \sin\theta)^{2} R^{\prime2}\right]\right\} \equiv G .
	\label{gaussap}
\end{equation}

\subsection{Contributions to $ \Delta $}% $ \Delta^{(0,0)}$,  $\Delta^{(2,0)}$ and $\Delta^{(0,2)}$}

We shall consider separately different spectral ranges.
For $ \omega\sim R_*^{-1} $, the argument of the exponential (\ref{gaussap}) is of order $ l_*^{-2} $, and we use the approximation $ G^{(s)}\simeq G^{(d)} \simeq 1$.
Therefore, $ \Delta^{(0,0)}$  gives the same result of the envelope approximation, $ \Delta_\mathrm{env} $. 
In this case the term $ \Delta^{(1,1)}  $ will be suppressed 
with respect to the other contributions. 
This can be seen, e.g., by integrating by parts the derivatives $ \partial_c $  in Eq.~(\ref{Pi11P}). Indeed, 
these derivatives only produce factors of order 1. 
In particular, the derivatives of $e^{i\omega\hat{n}\cdot\mathbf{s}} $
produce factors of order $ \omega R_* \sim  1$ (which will be important for higher frequencies). 
Since $\bar P \sim 1$, the contribution (\ref{Pi11P}) is suppressed by the factor $ l_* ^{-2} $. 
Besides the function $ \bar P(c) $ oscillates with frequency $ l_* $, which also causes a suppression of the integral.
Hence, in this frequency scale the spectrum is given by
\begin{equation}
	\Delta=\Delta^{(0,0)}+\Delta^{(2,0)}+\Delta^{(0,2)} =(1-2a^2)\Pi_\mathrm{env} .
	\label{Pi002002final}
\end{equation}

Let us now consider the case $ \omega\sim l_*/R_* $.  	
In this case the range of variation in $ R,R' $ of the Gaussian (\ref{gaussap}) becomes of order $ R_* $, so the presence of this function in the integrand begins to be noticeable.
Nevertheless, the function $ e^{-I_{\mathrm{tot}}} $ falls more quickly, since we have, roughly, 
$ G \sim e^{-a^2(R/R_*)^2}$ and $e^{-I_{\mathrm{tot}}} \sim e^{-I(t)}\sim e^{-\frac{4\pi}{3}(R/R_*)^3}$.
Therefore, the contribution (\ref{Delta00}) will not depart significantly from the result of the envelope approximation until higher $ \omega $, 
where the function $ G\sim \exp[{-(a\omega R_*/l_*)^2(R/R_*)^2}] $ will reduce the effective support for the integration.
We shall then assume that, at $ \omega \sim l_*/R_* $, the contribution $ \Delta^{(0,0)} $ is still roughly given by the envelope approximation, and concentrate
in the contribution $ \Pi^{(1,1)} $, which has a different behavior.

In the case $ \omega\sim k_*\sim l_*/R_* $,
if we integrate by parts  the second line of Eq.~(\ref{Pi11P}) there are terms which are not suppressed, namely, those containing 
derivatives of the exponential $ e^{i\omega\hat{n}\cdot\mathbf{s}} $. In fact, the dominant term is the one containing two derivatives, $ \partial_c^2 e^{i\omega\hat{n}\cdot\mathbf{s}} $.
Considering only this contribution, we obtain
\begin{multline}
	\Delta^{(1,1)}=- \frac{a^2v^2\omega^{5}}{3\pi R_*^{5}l_*^2}
\int_{0}^{\infty}dR' \int_{0}^{R'}dR \cos(v^{-1}\omega R_-) \left(RR'\right)^3 R^{\prime 2}
\\   \times
\int_{-1}^{+1} dc_\chi\int_0^{2\pi} d\phi
\int_{-1}^{+1}dc \,e^{-I_{\mathrm{tot}}} \, G \, e^{i\omega\hat{n}\cdot\mathbf{s}} 
 C \left(c_n - s_nc_\phi \frac{\cos \theta}{\sin\theta} \right)^2 \bar P(c),
 \label{Delta11}
\end{multline}
The  function $ e^{-I_{\mathrm{tot}}} $ sets an effective integration range for $ R $ and $ R' $ of order $R_*$.
For $ \omega \sim l_*/R_* $, the function $ G $ is smooth in this range, 
while the functions $ \cos(v^{-1}\omega R_-) $ and $ e^{i\omega \hat{n}\cdot \mathbf{s}} $ have strong oscillations.
Similarly, according to Eq.~(\ref{aproxlgrande}), the function $ \bar P (\cos\theta) $ oscillates with a very high frequency  $ l_*$.
We shall avoid the difficult oscillatory integrals by using approximations such as the stationary phase approximation and integrating by parts recursively to obtain an expansion in powers of $ \omega^{-1} $. The calculation is described in appendix \ref{apfaseestac}. We obtain
\begin{equation}
	\Delta^{(1,1)} (R_*\omega)=\frac{4a^{2}l_{*}^{3}}{3(R_*\omega)^{5}}\exp\left(-\frac{4\pi l_{*}^{3}}{3(R_*\omega)^{3}}\right). 
	\label{Delta11final}
\end{equation} 
This estimation is in principle only valid for $ R_*\omega \sim l_* $, but it falls quickly  from its peak at $ \omega_p= (4\pi/5)^{1/3} l_*/R_*$.
Around the maximum, the error of the approximation is of order $ 1/\sqrt{l_*} $.

\subsection{The spectrum}

For $ \omega\sim R_* $ the GW spectrum is given by Eq.~(\ref{Pi002002final}). 
To estimate the spectrum from the envelope approximation we shall use the results of Ref.~\cite{jt17}. 
That calculation was done for an exponential nucleation rate $ \Gamma=\Gamma_* e^{\beta(t-t_*)} $, while we are considering a simultaneous nucleation.  
Nevertheless, the two models have been compared by relating their characteristic length scales (see, e.g., \cite{gwlisa,w16,chw18}). Indeed, for the exponential rate the GW spectrum is determined by the parameters $ v $ and $ \beta $. For this model, the final bubble separation is given by $ R_*=(8\pi)^{1/3}v/\beta $, and we shall use this relation to write $ \beta $ as a function of $ R_* $. 
In Ref.~\cite{jt17}, a dimensionless spectrum $ \Delta_\beta $ was used. 
The definition of $ \Delta_\beta $ is similar to that of Eq.~(\ref{defDelta}), with $ R_* $ replaced by $ \beta^{-1} $. 
Hence, we have the relation 
$ \Delta(R_*\omega) = (R_* \beta)^{-2}\Delta_\beta(R_*\omega/(R_*\beta)) $, with a  conversion factor $ R_*\beta=(8\pi)^{1/3}v $. 
A fitting formula was provided in \cite{jt17} for the spectrum, 
\begin{equation}
	\Delta_\beta = {\Delta_p}\left[{c_l \left(\frac{\omega_p}{\omega}\right)^{3} + (1-c_l-c_h)\frac{\omega_p}{\omega}+c_h\frac{\omega}{\omega_p}}\right]^{-1},
\end{equation}
with $ c_l=0.064 $, $ c_h=0.48 $,  and
\begin{equation}
	 \Delta_p=0.48 v^3 /(1+5.3v^2+5v^4), \quad \omega_p/\beta= 2\pi \times 0.35/(1+0.069v+0.69v^4).
\end{equation}

\begin{figure}[tb]
	\centering
	\includegraphics[height=4.5cm]{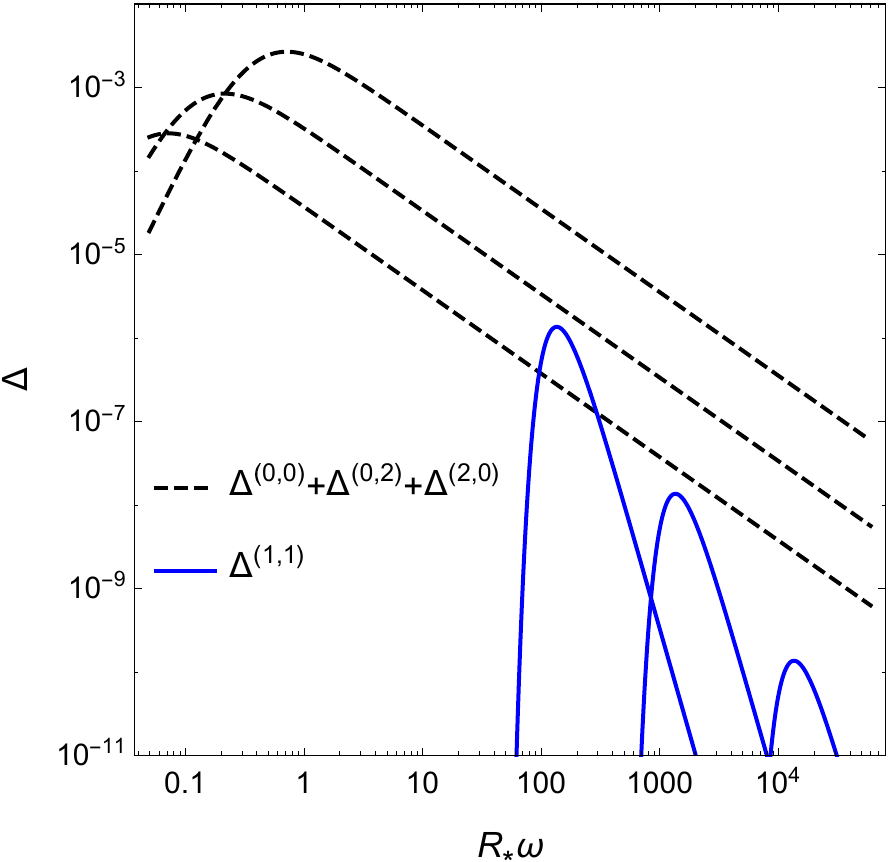}  \includegraphics[height=4.5cm]{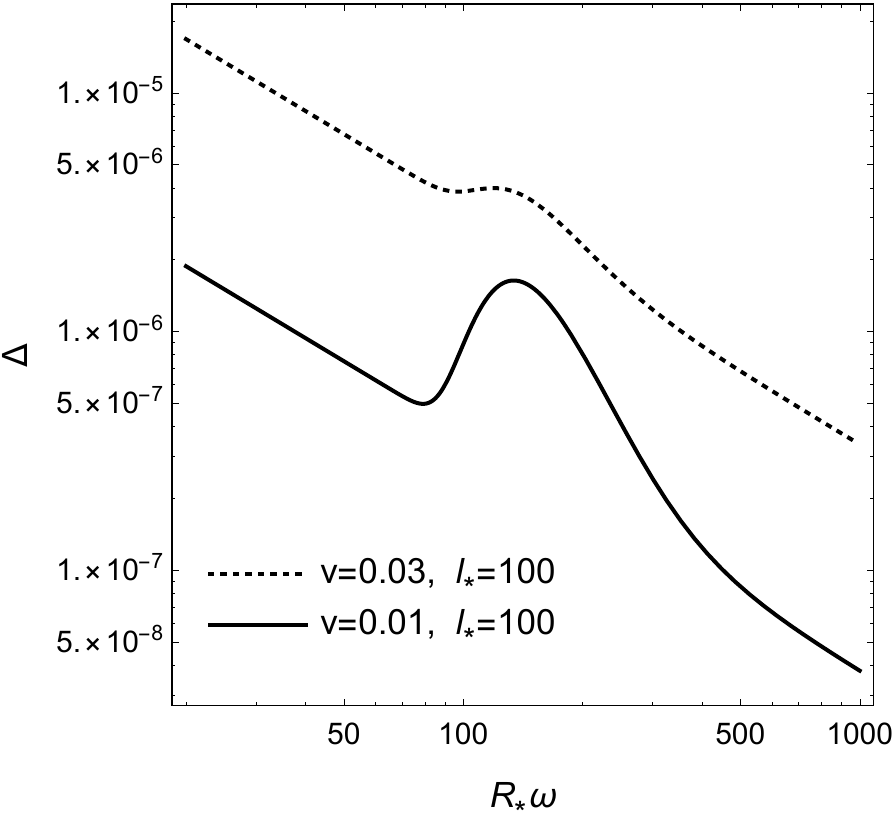} \includegraphics[height=4.5cm]{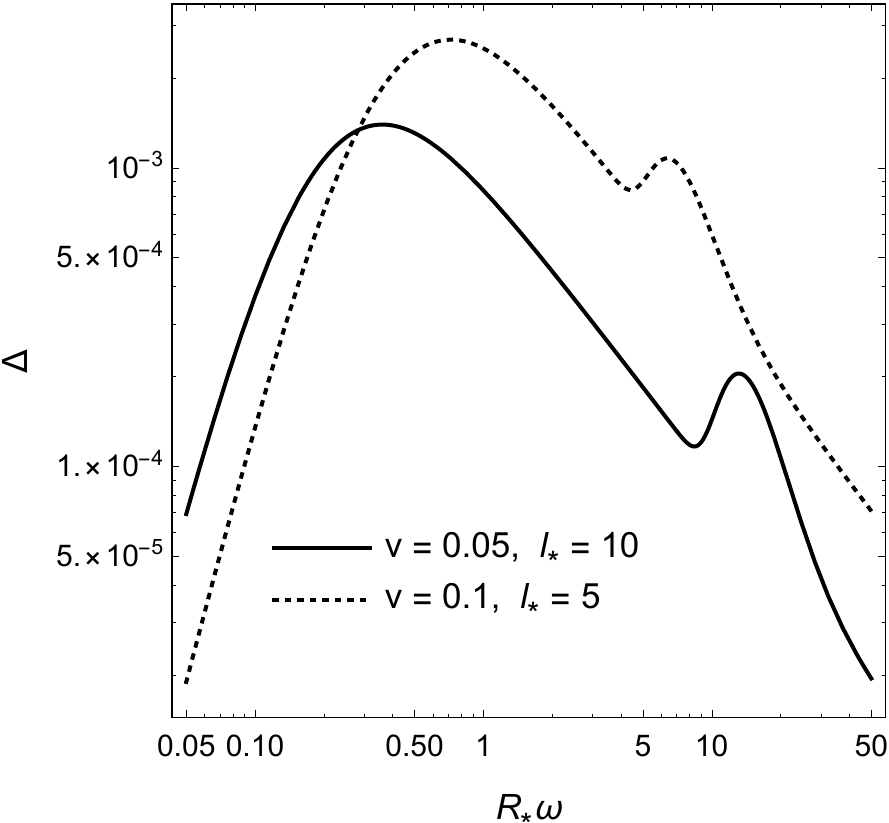}
	\caption{Left panel: The contribution $ \Delta^{(0,0)}+\Delta^{(2,0)}+\Delta^{(0,2)} $ for different velocities (from top to bottom, $ v=0.1,0.03,0.01 $), 
		and the contribution $ \Delta^{(1,1)} $ for different values of $ l_* $ (from top to bottom, $ l_*=100,1000,10000 $).
		Central panel: The sum of these contributions for two sets of the parameters. Right panel: Extrapolation to smaller values of $ l_* $.}
	\label{figcontrib}
\end{figure}

In the left panel of Fig.~\ref{figcontrib} we show the curves of  $(1-2a^2) \Delta_{\mathrm{env}} $ for a few velocities (dashed lines). All the curves correspond to $ a=0.5 $.
For a given value of $ l_* $, we have argued that this approximation for the contribution  $ \Delta^{(0,0)}+\Delta^{(2,0)}+\Delta^{(0,2)} $ should be valid up to $ w R_*\sim l_* $.
The contribution $ \Delta^{(1,1)} $, given by Eq.~(\ref{Delta11final}) is also shown in the left panel of  Fig.~\ref{figcontrib} for a few values of $ l_* $.
We see that the relative amplitude of the two contributions depends on the values of $ v $ and $ l_* $. 
In the central panel we show the sum of these contributions for two combinations of these parameters. The bump at $ \omega\sim l_*/R_* $ will be more pronounced for lower wall velocities and for smaller $ l_* $. 

We remark that our perturbative calculation has severely limited the intensity of the effect, since we needed to consider wall deformations with amplitude 
$ \zeta \sim a R_*/l_* $, with $ a<1 $. For $ a>1 $ the effect will be certainly stronger.
Besides, we used the approximation $ l_* \gg 1$ to neglect several terms and we approximated the integrals to lowest order in $ 1/l_* $. 
In particular, the latter have errors of order $ l_*^{-1/2} $,  so the value $ l_*=100 $ is already in the limit of validity of our estimations.
However, the only physical restriction on the deformation scale is $ k_*^{-1} <R_*$, i.e.,  $ l_* >1$.
Let us thus end with a speculation on the effect of deformations on a scale which is closer to the bubble size. 
For that aim we just extrapolate  our results to $ l_*=10 $ and $ l_*=5 $ in the right panel of Fig.~\ref{figcontrib}.

\section{Conclusions}

\label{conclu}

We have discussed a general method for calculating the gravitational waves generated
by the motion of thin bubble walls or thin fluid shells in a cosmological phase transition.
The main difference with previous approaches is in the calculation of the correlator of the
energy-momentum tensor $\left\langle T_{ij}(t,\mathbf{x})T_{kl}(t',\mathbf{x}')\right\rangle $,
although our approach takes ingredients from previous works \cite{kt93,jt17}.
In the technique introduced in Ref.~\cite{jt17}, two arbitrary space-time
points $t,\mathbf{x}$ and $t',\mathbf{x}'$ are considered, and then
the probability that bubbles are nucleated in the ``past light cones''
of these points is analyzed in order to determine whether a thin wall
is present at each of the events. In contrast, we directly consider
points on the bubble walls, which we assume from the beginning to
be infinitely thin, and we write $T_{ij}$ as a sum over bubbles,
as done in the approach introduced in Ref.~\cite{kt93}. On the other
hand, in the latter work the phase transition is simulated by nucleating
bubbles in a certain volume and following their evolution, while we
calculate the correlator statistically from the nucleation rate $\Gamma(t)$,
like in Ref.~\cite{jt17}.

We have exemplified the method by applying it to the bubble-collision mechanism in the envelope approximation and
to the bulk flow model.
The fact that we follow the motion of points on bubble walls makes
our method geometrically clear, while the statistical treatment facilitates
analytic calculations. 
Thus, for these cases  we have seen that some steps of the derivations
are much simpler than in other approaches. For instance, the decomposition
of $T_{ij}$ into a sum of individual bubble contributions gives naturally
a single-bubble contribution and a two-bubble contribution to $\langle T_{ij}T_{kl}\rangle$.
Furthermore, since we consider from the beginning only the uncollided
parts of a bubble surface, the breaking of the spherical symmetry
in the single-bubble case is also clear. Our method also allowed us
to arrive more directly to the general expressions for the bulk flow
model than it was done in the original work \cite{jt19} with the
approach of Ref.~\cite{jt17}, and we have also discussed alternative
expressions.

We have also considered the case of bubble walls which are deformed
from the spherical configuration. 
In this case, our general expression
for the GW spectrum contains the correlator of the surface deformation at
two points on a wall, $\left\langle \zeta(t,\hat{r})\zeta(t',\hat{r}')\right\rangle $,
which depends on the specific dynamics of the deformations. This result
can be applied, e.g., to the case of inhomogeneous wall velocities due to inhomogeneous reheating during the phase transition or to
the case of wall corrugations arising from hydrodynamic instabilities of deflagrations.
For small enough deformations, 
the general expressions can be expanded perturbatively, which simplifies considerably the calculations.
The lowest order is the result of the envelope approximation. 
However, the corrections introduce a suppression factor to this component,
as well as a contribution which becomes relevant at frequencies associated to the deformation dynamics.

As a simple application we have modeled a wall deformation spectrum
which peaks at a single length scale $ k_*^{-1} $.
This introduces a peak in the GW spectrum at $ \omega\sim k_* $, which is always at a higher scale than that corresponding to the bubble size, $ R_*^{-1} $.
The perturbative expansion requires considering $ \zeta\lesssim R_* $ as well as  $ k_*\zeta \lesssim 1$, 
which implies that, within this approximation, it is impossible to obtain an important effect for very large $ k_* $.
We do have a clear difference (a second peak) from the GW spectrum of the bubble-collision mechanism for $ k_*\lesssim 100/R_* $.
We remark that this is a conservative estimate, and the effect may be quite larger. 
The wall deformations may range from scales of order $  R_*$ (from bubble interactions) to several orders of magnitude smaller (from instabilities).
In the first case, the spectrum will have a bump close to the peak of the traditional GW generation mechanisms (bubble collisions, sound waves, turbulence).
For the electroweak phase transition, this effect could be detected by LISA. 
In the case of much smaller deformations, there will be a peak at $ \omega\gg R_*^{-1} $, which could be observed by other
future detectors such as BBO \cite{BBO},  DECIGO \cite{DECIGO}, ET \cite{ET}, or AEDGE \cite{AEDGE}. 

\section*{Acknowledgments}

This work was supported by CONICET grant PIP 11220130100172 
and Universidad Nacional de Mar del Plata, grant EXA999/20.

\appendix

\section{Gravitational waves from a stochastic source}

\label{approaches}

Consider the gravitational waves emitted by a volume $V$ in a direction
$\hat{n}$. If this radiation is observed in the wave zone, the total
energy per solid angle and per unit frequency interval is given by
\cite{WeinbergGC} 
\begin{equation}
\frac{dE}{d\ln\omega d\Omega}=2G\omega^{3}\Lambda_{ij,kl}(\hat{n})T_{ij}(\omega,\hat{n})^{*}T_{kl}(\omega,\hat{n}),\label{weinberg}
\end{equation}
where 
\begin{equation}
T_{ij}(\omega,\hat{n})=\int_{-\infty}^{\infty}\frac{dt}{2\pi}e^{i\omega t}\int_{V}d^{3}xe^{-i\omega\hat{n}\cdot\mathbf{x}}T_{ij}(t,\mathbf{x}).\label{Tkom}
\end{equation}
In Refs.~\cite{kt93,hk08}, $T_{ij}$ is written as a sum over bubbles
and the phase transition is simulated with a certain number of bubbles.
We shall also use the explicit decomposition $T_{ij}=\sum_{n}T_{ij}^{(n)}$,
but, since the contribution of many bubbles makes the phase transition
a stochastic source, we shall consider the average on a statistical
ensemble. Due to the homogeneity and isotropy of this source, the
energy radiated in direction $\hat{n}$ does not actually depend on
$\hat{n}$. Therefore, integrating the angular variables in Eq.~(\ref{weinberg})
only gives a factor of $4\pi$. Thus, considering a large volume $V$,
we may write 
\begin{equation}
\frac{d\rho_{GW}}{d\ln\omega}=\frac{4\pi}{V}\left\langle \frac{dE}{d\ln\omega d\Omega}\right\rangle ,
\end{equation}
 and using  Eqs.~(\ref{weinberg}-\ref{Tkom}) we obtain 
\begin{equation}
\frac{d\rho_{GW}}{d\ln\omega}=\frac{2G\omega^{3}}{\pi}\int_{-\infty}^{\infty}dt\int_{-\infty}^{\infty}dt'e^{i\omega(t-t')}\Pi(t,t',\omega),\label{rhogwTF}
\end{equation}
 where $\Pi$ is the quantity defined in Eq.~(\ref{defPi}), 
\begin{equation}
\Pi=\frac{1}{V}\int_{V}d^{3}x\int_{V}d^{3}x'e^{-i\omega\hat{n}\cdot(\mathbf{x}-\mathbf{x}')}\left\langle \Lambda_{ij,kl}(\hat{n})T_{ij}(t,\mathbf{x})T_{kl}(t',\mathbf{x}')\right\rangle .\label{formulaPi}
\end{equation}

It is convenient to find an alternative expression, in which the times
$t$ and $t'$ have a definite time ordering. For that aim, we shall
first show that $\Pi$ is real, does not depend on $\hat{n}$, and
satisfies $\Pi(t,t',\omega)=\Pi(t',t,\omega)$. These properties follow from those of the quantity
\begin{equation}
\langle\Lambda TT\rangle\equiv\left\langle \Lambda_{ij,kl}(\hat{n})T_{ij}(t,\mathbf{x})T_{kl}(t',\mathbf{x}')\right\rangle .
\end{equation}
By translation symmetry, this quantity depends only on the difference $\boldsymbol{r}=\mathbf{x}-\mathbf{x}'$.
Hence, one of the integrals in Eq.~(\ref{formulaPi}) gives a factor of $ V $,
\begin{equation}
\Pi =\int_{V}d^{3}r\, e^{-i\omega\hat{n}\cdot\mathbf{r}}\langle\Lambda TT\rangle(t,t',\mathbf{r})\label{Pi1int}
\end{equation}
By rotation symmetry, $ \langle\Lambda TT\rangle $ only depends on $ r $ and on 
$ \cos\chi=\hat{r} \cdot\hat{n}$. For very large $V$,   we have
\begin{equation}
	\Pi	=2\pi\int_{0}^{\infty}drr^{2}\int_{0}^{\pi}d\chi\sin\chi e^{-i\omega r\cos\chi}\langle\Lambda TT\rangle(t,t',r,\cos\chi).
\end{equation}
The dependence on $\hat{n}$ disappears upon integration on $\chi$. 

Using the properties $\Lambda_{ij,kl}(\hat{n})=\Lambda_{kl,ij}(\hat{n})$ and $\Lambda_{ij,kl}(-\hat{n})=\Lambda_{ij,kl}(\hat{n})$,
we have 
\begin{equation}
\langle\Lambda TT\rangle(t,t',r,\hat{n}\cdot\hat{r})=\langle\Lambda TT\rangle(t',t,r,-\hat{n}\cdot\hat{r})=\langle\Lambda TT\rangle(t',t,r,\hat{n}\cdot\hat{r}). \label{ttptpt}
\end{equation} 
This implies that $\Pi(t,t',\omega)=\Pi(t',t,\omega)$. Besides, $\langle\Lambda TT\rangle$
is real, and Eq.~(\ref{ttptpt}) shows that it is even in $\cos\chi$,
so we can write 
\begin{equation}
\Pi=2\pi\int_{0}^{\infty}drr^{2}\int_{-1}^{+1}du\cos(\omega ru)\langle\Lambda TT\rangle(t,t',r,u),
\end{equation}
which is real. Since $\rho_{GW}$ and $\Pi$ are real, we may take
the real part of the expression (\ref{rhogwTF}), so that the exponential
becomes a cosine. Then, using the symmetry of $\Pi$ under $t\leftrightarrow t'$
to write $\int_{-\infty}^{+\infty}dt'=2\int_{t}^{\infty}dt'$, we
obtain Eq.~(\ref{rhogw}),
\begin{equation}
\frac{d\rho_{GW}}{d\ln\omega}=\frac{4G\omega^{3}}{\pi}\int_{-\infty}^{\infty}dt\int_{t}^{\infty}dt'\cos[\omega(t-t')]\,\Pi(t,t',\omega).\label{rhogwapen}
\end{equation}

Since our starting point, Eq.~(\ref{weinberg}), was the same as
in Refs.~\cite{kt93,hk08}, and the contribution of many bubbles
makes the phase transition a stochastic source, our treatment should
be equivalent. It is not difficult to see also the equivalence
of Eq.~(\ref{rhogwapen}) with the one used in Refs.~\cite{jt17,cds08}.
There, the evolution of the metric perturbations $h_{ij}(t,\mathbf{k})$
is solved with the Green function method, which relates $h_{ij}$
with $\Pi_{ij}(t,\mathbf{k})=\Lambda_{ij,kl}T_{kl}(t,\mathbf{k})$.
Then, the energy density of the stochastic background of GWs is obtained
from $\langle\dot{h}_{ij}(t,\mathbf{k})\dot{h}_{ij}^{*}(t,\mathbf{q})\rangle$.
In this approach, the quantity $\Pi$ is defined through the unequal-time
correlator 
\begin{equation}
\left\langle \Pi_{ij}(t,\mathbf{k})\Pi_{ij}^{*}(t',\mathbf{q})\right\rangle =(2\pi)^{3}\delta^{(3)}(\mathbf{k}-\mathbf{q})\Pi(t,t',k).\label{PiPidelta}
\end{equation}
Since
$
	T_{kl}(t,\mathbf{k})=\int d^{3}xe^{i\mathbf{k\cdot}\mathbf{x}}T_{kl}(t,\mathbf{x})\label{Tkl},
$
and taking into account the projector property of $ \Lambda_{ij,kl} $, namely,
$\Lambda_{ij,kl}\Lambda_{ij,mn}=\Lambda_{kl,mn}$,
we have 
\begin{equation}
\left\langle \Pi_{ij}(t,\mathbf{k})\Pi_{ij}^{*}(t',\mathbf{q})\right\rangle =\int d^{3}xe^{i\mathbf{k\cdot}\mathbf{x}}\int d^{3}x'e^{i\mathbf{q\cdot}\mathbf{x}'}\langle\Lambda TT\rangle(t,t',\mathbf{x}-\mathbf{x}').
\end{equation}
The translation invariance gives a delta function, and comparing with (\ref{PiPidelta})
we obtain
\begin{equation}
\Pi(t,t',k)=\int d^{3}re^{i\mathbf{k\cdot}\mathbf{r}}\langle\Lambda TT\rangle(t,t',\mathbf{r})
\end{equation}
 (cf.~Eqs.(18-19) of Ref.~\cite{jt17}). Writing $\mathbf{k}=\omega\hat{n}$, this expression essentially coincides with our Eq.~(\ref{Pi1int}).

\section{Averaging the direction of observation}

\label{angint}

In this appendix we calculate the angular average 
\begin{equation}
\frac{1}{4\pi}\int d\hat{n}\,e^{i\omega\hat{n}\cdot\mathbf{s}}\Lambda_{ij,kl}(\hat{n})\hat{r}_{i}\hat{r}_{j}\hat{r}'_{k}\hat{r}'_{l},
\end{equation}
and then the angular integrals in Eqs.~(\ref{Piscol}) and (\ref{Pidcol}).

\begin{figure}[tb]
\centering
\includegraphics[height=4cm]{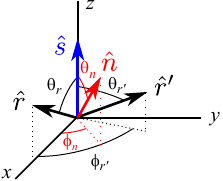} ~ ~ ~ \includegraphics[height=4cm]{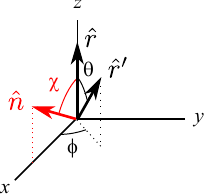}
\caption{The unit vectors  $\hat{r}$, $\hat{r}'$, $\hat{s}$, $\hat{n}$ and their angles for two different orientations of the axes.}
\label{figangles}
\end{figure}

The quantity $\Lambda_{ij,kl}(\hat{n})\hat{r}_{i}\hat{r}_{j}\hat{r}'_{k}\hat{r}'_{l}$,
which is given by Eq.~(\ref{Ltt0}), contains angles of $\hat{n}$
with $\hat{r}$ and $\hat{r}'$. To perform the integral over $\hat{n}$,
it is convenient to put the $z$ axis in the direction of the vector
$\mathbf{s}$, so that the exponential $e^{i\omega\hat{n}\cdot\mathbf{s}}$
depends on a single angle. We also put the vector $\hat{r}$ in
the $xz$ plane (see Fig.~\ref{figangles}), which simplifies a little
the expressions. 
We thus write $\hat{s}=(0,0,1)$, $\hat{r}=(s_{r},0,c_{r})$, with
$c_{r}\equiv\cos\theta_{r}$ and $s_{r}^{2}=1-c_{r}^{2}$, $\hat{r}'=(s_{r'}\cos\phi_{r'},s_{r'}\sin\phi_{r'},c_{r'})$,
with $c_{r'}=\cos\theta_{r'}$ and $s_{r'}^{2}=1-c_{r'}^{2}$, and
$\hat{n}=(s_{n}\cos\phi_{n},s_{n}\sin\phi_{n},c_{n})$, with $c_{n}=\cos\theta_{n}$
and $s_{n}^{2}=1-c_{n}^{2}$. Thus, we have $\hat{n}\cdot\hat{s}=c_{n}$,
$d\hat{n}=dc_{n}d\phi_{n}$ and the exponential becomes $e^{i\omega sc_{n}}$.
On the other hand, the quantity $\Lambda_{ij,kl}(\hat{n})\hat{r}_{i}\hat{r}_{j}\hat{r}'_{k}\hat{r}'_{l}$
contains the scalar products 
\begin{align}
\hat{n}\cdot\hat{r} & =s_{r}s_{n}\cos\phi_{n}+c_{r}c_{n}.\label{nr}\\
\hat{n}\cdot\hat{r}' & =s_{r'}s_{n}\cos\phi_{r'}\cos\phi_{n}+s_{r'}s_{n}\sin\phi_{r'}\sin\phi_{n}+c_{r'}c_{n}.\label{nrp}
\end{align}
Inserting Eqs.~(\ref{nr}-\ref{nrp}) in Eq.~(\ref{Ltt0}), the
expression becomes rather cumbersome. Nevertheless, the integral with
respect to $\phi_{n}$ is straightforward, since we only have a few
powers of $\cos\phi_{n}$ and $\sin\phi_{n}$. After integrating this
angle, we obtain a simple polynomial in $c_{n}$, of the form $A+Bc_{n}^{2}+Cc_{n}^{4}$.
The integrals $\int_{-1}^{1}dc_{n}e^{i\omega sc_{n}}c_{n}^{k}$ are
readily evaluated. The result can be written in terms of the spherical
Bessel functions (\ref{SBessel}). We obtain 
\begin{equation}
\frac{1}{4\pi}\int d\hat{n}\,e^{i\omega\hat{n}\cdot\mathbf{s}}\Lambda_{ij,kl}(\hat{n})\hat{r}_{i}\hat{r}_{j}\hat{r}'_{k}\hat{r}'_{l}=C_{0}j_{0}(\omega s)+C_{1}\frac{j_{1}(\omega s)}{\omega s}+C_{2}\frac{j_{2}(\omega s)}{(\omega s)^{2}}.\label{integnap}
\end{equation}
The coefficients $C_{i}$ depend on $c_{r}$, $c_{r'}$, $\phi_{r'}$,
and $\hat{r}\cdot\hat{r}'$. The latter two variables are related
by
\begin{equation}
\hat{r}\cdot\hat{r}'=s_{r}s_{r'}\cos\phi_{r'}+c_{r}c_{r'}.\label{rrpfi}
\end{equation}
We can eliminate $\phi_{r'}$, and we obtain 
\begin{align}
C_{0} & =-\frac{1}{2}+\frac{c_{r}^{2}}{2}+\frac{c_{r'}^{2}}{2}+\frac{c_{r}^{2}c_{r'}^{2}}{2}-2c_{r}c_{r'}\left(\hat{r}\cdot\hat{r}'\right)+\left(\hat{r}\cdot\hat{r}'\right)^{2},\nonumber \\
C_{1} & =1-c_{r}^{2}-c_{r'}^{2}-5c_{r}^{2}c_{r'}^{2}+8c_{r}c_{r'}\left(\hat{r}\cdot\hat{r}'\right)-2\left(\hat{r}\cdot\hat{r}'\right)^{2},\label{c1ap}\\
C_{2} & =\frac{1}{2}-\frac{5}{2}c_{r}^{2}-\frac{5}{2}c_{r'}^{2}+\frac{35}{2}c_{r}^{2}c_{r'}^{2}-10c_{r}c_{r'}\left(\hat{r}\cdot\hat{r}'\right)+\left(\hat{r}\cdot\hat{r}'\right)^{2}.\nonumber 
\end{align}
Using $c_{r}=\hat{r}\cdot\hat{s}$, $c_{r'}=\hat{r}'\cdot\hat{s}$,
we obtain the expressions given in Eqs.~(\ref{C0})-(\ref{C2}).

For the single-bubble case, $\hat{s}$ is related to $\hat{r}$ and
$\hat{r}'$ through $s\hat{s}=R'\hat{r}'-R\hat{r}$, so we have 
\begin{equation}
\hat{r}\cdot\hat{s}=(R'\hat{r}\cdot\hat{r}'-R)/s,\quad\hat{r}'\cdot\hat{s}=(R'-R\hat{r}\cdot\hat{r}')/s,
\end{equation}
and Eq.~(\ref{integnap}) depends on a single cosine, which is related
to $s$ by
\begin{equation}
\hat{r}\cdot\hat{r}'=(R^{\prime2}+R^{2}-s^{2})/2RR'.\label{crrp}
\end{equation}
Thus, we obtain

\begin{equation}
\frac{1}{4\pi}\int d\hat{n}\,e^{i\omega\hat{n}\cdot\mathbf{s}}\left(\Lambda_{ij,kl}\hat{r}_{i}\hat{r}_{j}\hat{r}'_{k}\hat{r}'_{l}\right)=\frac{1}{32R^{\prime2}R^{2}s^{4}}\sum_{i=0}^{2}P_{i}\frac{j_{i}(\omega s)}{(\omega s)^{i}},
\end{equation}
with
\begin{align}
P_{0}= & \,\left[(R'-R)^{2}-s^{2}\right]^{2}\left[(R'+R)^{2}-s^{2}\right]^{2},\\
P_{1}= & \,2\left[(R'-R)^{2}-s^{2}\right]\left[(R'+R)^{2}-s^{2}\right]\left[3s^{4}+2s^{2}(R^{\prime2}+R^{2})-5(R^{2}-R^{\prime2})^{2}\right],\\
P_{2}= & \,3s^{8}+4s^{6}(R^{\prime2}+R^{2})+6s^{4}(3R^{\prime4}+2R^{\prime2}R^{2}+3R^{4})\nonumber \\
 & \,-60s^{2}(R^{2}-R^{\prime2})^{2}(R^{\prime2}+R^{2})+35(R^{2}-R^{\prime2})^{4}.
\end{align}
Since this result depends only on the cosine $c=\hat{r}\cdot\hat{r}'$,
the angular integrals in Eq.~(\ref{Piscol}) can be written as $\int d\hat{r}\int d\hat{r}'=8\pi^2\int_{-1}^{+1}dc=8\pi^{2}\int_{R_{-}}^{R_{+}}\frac{sds}{RR'}$.

For the two-bubble case, the vector $\mathbf{s}$ is independent of
$\hat{r}$ and $\hat{r}'$, and the angular integrals $\int d\hat{s}\int d\hat{r}\int d\hat{r}'$
in Eq.~(\ref{bracketd}) are straightforward. We only need to take
into account the restrictions (\ref{cM})-(\ref{cMp}), $-c_{M}\leq\hat{r}\cdot\hat{s}\leq1$
and $-1\leq\hat{r}'\cdot\hat{s}\leq c_{M}^{\prime}$. For the integral
over $\hat{r}'$ we may put $\hat{s}$ in the $z$ axis and $\hat{r}$
in the $xz$ plane, like in Fig.~\ref{figangles} (where the variable
$\hat{n}$ is already integrated). Thus, the quantities $C_{i}$ in
Eqs.~(\ref{c1ap}) depend on $\phi_{r'}$ through Eq.~(\ref{rrpfi}),
and the integration on this variable is trivial. We obtain 
\begin{equation}
\int_{0}^{2\pi}d\phi_{r'}C_{0}=0,\quad\int_{0}^{2\pi}d\phi_{r'}C_{1}=0,\quad\int_{0}^{2\pi}d\phi_{r'}C_{2}=2\pi\left(1-3c_{r}^{2}\right)\left(1-3c_{r'}^{2}\right).
\end{equation}
The result depends only on $c_{r}=\hat{r}\cdot\hat{s}$, $c_{r'}=\hat{r}\cdot\hat{s}$,
and the integral on the azimuth $\phi_{r}$ gives a factor of $2\pi$.
The integrals on the polar angles are also trivial and we only have
to take into account the conditions $-c_{M}\leq c_{r}\leq1$ and $-1\leq c_{r'}\leq c_{M}^{\prime}$.
If both limits $c_{M},c_{M}'$ are in the range $[-1,1]$, we have
\begin{equation}
\int_{-c_{M}}^{1}dc_{r}\left(1-3c_{r}^{2}\right)\int_{-1}^{c_{M}^{\prime}}dc_{r'}\left(1-3c_{r'}^{2}\right)=(c_{M}-c_{M}^{3})(c_{M}^{\prime}-c_{M}^{\prime3}).
\end{equation}
Otherwise, the result vanishes. Finally, the integral over $\hat{s}$
gives a factor of $4\pi$.

\section{Averaging the deformations}
\label{apave}

We consider the average 
\begin{equation}
	\left\langle e^{i(a\zeta+b\zeta')}\right\rangle \equiv G(a,b).  \label{expmedia}
\end{equation}
%where $a=-\omega\hat{n}\cdot\hat{r}$, $b=\omega\hat{n}\cdot\hat{r}'$.
The function $G$ is the characteristic function
for the two variables $\zeta$, $\zeta'$. We shall assume a normal distribution
%These variables are not independent, but it is generally valid to
%assume that they are jointly normal\footnote{See, e.g., \cite{lemons}.},
%so the variable $X=a\zeta+b\zeta'$ is normal. Thus, assuming also
with
$\langle\zeta\rangle=\langle\zeta'\rangle=0$. Therefore,  we have\footnote{See, e.g., \cite{vankampen}.} 
\begin{equation}
	G(a,b)=
	\exp\left[-\frac{a^{2}\left\langle \zeta^{2}\right\rangle +b^{2}\left\langle \zeta^{\prime2}\right\rangle +2ab\left\langle \zeta\zeta'\right\rangle }{2}\right].
	\label{charfunc}
\end{equation}
%For the two-bubble contribution, we have 
%$\left\langle \zeta\zeta'\right\rangle =\left\langle \zeta\right\rangle \left\langle \zeta'\right\rangle =0$,
%which gives $G=\exp\left[-\frac{1}{2}\left(a^{2}\left\langle \zeta^{2}\right\rangle +b^{2}\left\langle 
%\zeta^{\prime2}\right\rangle \right)\right]$.
Any expectation value can be derived from the characteristic functional for the entire stochastic 
process\footnote{Namely, $G[a]= \left\langle \exp \left[i\int a(\hat r,t)\zeta(\hat r,t) d\hat r dt\right] \right\rangle$.}.
However, for our purposes it suffices with the characteristic function for a few points. 
%For instance, 
%applying $-i\partial/\partial a$ to Eq.~(\ref{expmedia}), we obtain
%\begin{equation}
%	\left\langle \zeta e^{i(a\zeta+b\zeta')}\right\rangle =i\left(a\left\langle \zeta^{2}\right\rangle +b\left\langle 
%   \zeta\zeta'\right\rangle \right)\varphi_{\zeta,\zeta'}.\label{zetamedio}
%\end{equation}
For instance, applying $-i\partial/\partial x_{i}$ to 
(\ref{expmedia})-(\ref{charfunc})\footnote{Notice that, for these derivations, we may consider the variables $a,b$ as independent of $x_{i}$ or $x_k'$.
Later we may evaluate the result at $a=-\omega\hat{n}\cdot\hat{r}$, $b=\omega\hat{n}\cdot\hat{r}'$.} 
and taking into account that, by symmetry, $\left\langle \zeta^{2}\right\rangle $
does not depend on the point on the bubble wall, and $\left\langle \zeta\partial_{i}\zeta\right\rangle =0$, we obtain
\begin{equation}
	\left\langle \partial_{i}\zeta e^{i(a\zeta+b\zeta')}\right\rangle =ib\left\langle \partial_{i}\zeta\zeta'\right\rangle G
	 %\quad 
	 %\left\langle \partial_{k}\zeta'e^{i(a\zeta+b\zeta')}\right\rangle =ia\left\langle \zeta\partial_{k}\zeta'\right\rangle G
	 .
	 \label{dizmedio}
\end{equation}
Applying  $-i\partial/\partial x_{k}'$ to the latter, we obtain
\begin{equation}
	\left\langle \partial_{i}\zeta\partial_{k}\zeta'e^{i(a\zeta+b\zeta')}\right\rangle =\left[\left\langle \partial_{i}\zeta\partial_{k}\zeta'\right\rangle -ab 
	\left\langle \zeta'\partial_{i}\zeta\right\rangle\left\langle \zeta\partial_{k}\zeta'\right\rangle \right]G.
\end{equation}
If we consider a third point $ \zeta''=\zeta(\hat{r}'',t'') $, the generating function $ G(a,b,c) $ is the trivial generalization of Eqs.~(\ref{expmedia})-(\ref{charfunc}). We thus obtain the generalization of  Eq.~(\ref{dizmedio}),
\begin{equation}
	\left\langle \partial_{i}\zeta e^{i(a\zeta+b\zeta'+c\zeta'')}\right\rangle =i\left[b\left\langle \zeta'\partial_{i}\zeta\right\rangle +c\left\langle \zeta''\partial_{i}\zeta\right\rangle \right]G(a,b,c),
\end{equation}
and, applying  $-i\partial/\partial x_{j}''$, we have
\begin{equation}
\left\langle \partial_{i}\zeta\partial_{j}\zeta''e^{i(a\zeta+b\zeta'+c\zeta'')}\right\rangle =\left\langle \partial_{j}\zeta''\partial_{i}\zeta\right\rangle \varphi-\left(b\left\langle \zeta'\partial_{i}\zeta\right\rangle +c\left\langle \zeta''\partial_{i}\zeta\right\rangle \right)\left(a\left\langle \zeta\partial_{j}\zeta''\right\rangle +b\left\langle \zeta'\partial_{j}\zeta''\right\rangle \right)G.
\end{equation}
Evaluating in $ c=0 $ and $ \zeta''=\zeta $, we obtain
\begin{equation}
	\left\langle \partial_{i}\zeta\partial_{j}\zeta e^{i(a\zeta+b\zeta')}\right\rangle =\left[\left\langle \partial_{i}\zeta\partial_{j}\zeta\right\rangle -b^{2}\left\langle \zeta'\partial_{i}\zeta\right\rangle \left\langle \zeta'\partial_{j}\zeta\right\rangle \right]G.
\end{equation}

\section{Approximations for $ \Delta^{(1,1)} $}
\label{apfaseestac}

We shall estimate the integrals in Eq.~(\ref{Delta11}), which we write in the form 
\begin{multline}
	\Delta^{(1,1)}=- \frac{a^2v^2\omega^{5}}{6\pi R_*^{5}l_*^2}
	\int_{0}^{\infty}dR_+ \int_{0}^{R_+}dR_-\cos(v^{-1}\omega R_-) \left(\frac{R_+^2-R_-^2}{4}\right)^3 \left(\frac{R_++R_-}{2}\right)^2
	\\   \times
	\int_{-1}^{+1} dc_\chi\int_0^{2\pi} d\phi
	\int_{-1}^{+1}dc \,e^{-I_{\mathrm{tot}}(s,t,t')} \, G \, e^{i\omega\hat{n}\cdot\mathbf{s}} 
	\left(\sin\theta c_{\chi}-\cos\theta s_{\chi}c_{\phi}\right)^{2}\frac{C}{\sin^{2}\theta} \bar P(c),
	\label{Delta11apen}
\end{multline}
where 
\begin{equation}
	G=\exp\left\{ -\frac{1}{2}\frac{\omega^2 a^2}{l_{*}^2}
	\left[c_{\chi}^{2}\left(\frac{R_{+}-R_{-}}{2}\right)^{2}+(c_{\chi}c+s_{\chi}\cos\phi\sin\theta)^{2}\left(\frac{R_{+}+R_{-}}{2}\right)^{2}\right]\right\},
\end{equation}
\begin{equation}
	\frac{C}{\sin^{2}\theta}=2\left[2\left(s_{\chi}^{2}c_{\phi}^{2}-1\right)s_{\chi}^{2}c^{2}
	-\left(s_{\chi}^{2}c_{\phi}^{2}-c_{\chi}^{2}\right)c_{\chi}s_{\chi}c_{\phi}c\sin\theta+s_{\chi}^{2}\left(1-s_{\chi}^{2}c_{\phi}^{2}\right)\right]
\end{equation}
and 
\begin{equation}
	e^{i\omega\hat n\cdot\mathbf{s}}=\exp\left\{i\omega \left[(c_{\chi}c+s_{\chi}c_{\phi}\sin\theta-c_{\chi})\frac{R_{+}}{2}
+ (c_{\chi}c+s_{\chi}c_{\phi}\sin\theta+c_{\chi})\frac{R_-}{2}  \right]\right\}.
\label{eiwns}
\end{equation}

We first approximate the integral with respect to $ R_- $. We have an integral of the form 
\begin{equation}
I_{R_-}=\int_{0}^{R_{+}}dR_{-}\left(e^{i\omega y_{+}R_{-}}+e^{i\omega y_{-}R_{-}}\right)F(R_{-}),
\end{equation}
with $ y_{\pm}=v^{-1}+\frac{1}{2}\left(c_{\chi}\cos\theta+s_{\chi}c_{\phi}\sin\theta+c_{\chi}\right) $ and 
$ F(R_{-})=\left(\frac{R_{+}^{2}-R_{-}^{2}}{4}\right)^{3}\left(\frac{R_{+}+R_{-}}{2}\right)^{2}e^{-I_{\mathrm{tot}}}G $.
Integrating by parts twice, we obtain 
\begin{equation}
I_{R_-}=\frac{v^{2}}{i\omega}\frac{\psi}{1-\frac{v^{2}}{4}\psi^{2}}F(0)
-\frac{2v^{2}}{\omega^{2}}\frac{1+\frac{v^{2}}{4}\psi^{2}}{\left(1-\frac{v^{2}}{4}\psi^{2}\right)^{2}}\frac{\partial F}{\partial R_{-}}(0)
+\mathcal{O}\left(\frac{1}{\omega^{3}}\right),
\label{IRm}
\end{equation}
where
$ \psi=c_{\chi}\cos\theta+s_{\chi}c_{\phi}\sin\theta+c_{\chi} $. This variable will vanish upon angular integration.
This is why we kept the second order term in Eq.~(\ref{IRm}).
Denoting $ \bar{R}=R_{+}/2 $, we have $ F(0)=\bar{R}^{8}e^{-I_{\mathrm{tot}}}G $ and 
\begin{equation}
	F'(0)=\bar{R}^{7}\left\{ 1-\frac{1}{2}\frac{\omega^2 a^2}{l_{*}^2}\bar{R}^{2}\left[(c_{\chi}\cos\theta+s_{\chi}\cos\phi\sin\theta)^{2}-c_{\chi	}^{2}\right]\right\} e^{-I_{\mathrm{tot}}}G, 
\end{equation}
where now we have 
\begin{equation}
	 I_{\mathrm{tot}}=\frac{\pi\bar{R}^{3}}{6R_{*}^{3}}\left[8+6\sqrt{2(1-c)}-\sqrt{2}(1-c)^{3/2}\right] 
\end{equation}
and 
\begin{equation}
	G=\exp\left\{ -\frac{1}{2}\frac{\omega^2 a^2}{l_{*}^2}\left[c_{\chi}^{2}+(c_{\chi}\cos\theta+s_{\chi}\cos\phi\sin\theta)^{2}\right]\bar{R}^{2}\right\} 
\end{equation}

Now we will approximate the integral with respect to $ \phi $. We have an integral of the form 
\begin{equation}
	I_\phi = \int_{0}^{2\pi}d\phi e^{i\omega s_{\chi}\sin\theta\bar{R}\cos{\phi}}H(\phi),
\end{equation}
where $ H=\frac{C}{\sin^{2}\theta}\left(\sin\theta c_{\chi}-\cos\theta s_{\chi}c_{\phi}\right)^{2}I_{R_{-}} $.
For large $ \omega $, we can use the stationary phase approximation. Since the integrand is periodic in $ \phi $ we shift the integration interval to $ [-\pi/2,3\pi/2] $, where the cosine has stationary points at $ \phi =0$ and $ \phi =\pi $. We obtain
\begin{equation}
	I_\phi = e^{i\omega\bar{R}s_{\chi}\sin\theta}H_{-}+e^{-i\omega\bar{R}s_{\chi}\sin\theta}H_{+}, 
	\label{Iphi}
\end{equation}
where
\begin{equation}
	H_{-}=e^{-i\pi/4}\sqrt{\frac{2\pi}{\omega\bar{R}s_{\chi}\sin\theta}}H(0),\qquad H_{+}=e^{+i\pi/4}\sqrt{\frac{2\pi}{\omega\bar{R}s_{\chi}\sin\theta}}H(\pi).
\end{equation}

Next, we estimate the integral with respect to $ c_\chi=\cos \chi $. 
The exponentials in (\ref{Iphi}) combine with the remaining exponentials in (\ref{eiwns})
and give exponents of the form $ i\omega\bar R[\left(\cos(\chi\mp\theta)-\cos\chi\right)]$. 
For $ \chi $ and $ \theta $ in the interval $ [0,\pi] $, there is only one stationary point for each case, at $ \chi_\mp =\pi/2\pm\theta/2 $, and we obtain
\begin{align}
\int_{0}^{\pi}d\chi\sin\chi e^{i\omega\bar{R}\left[\cos(\chi\mp\theta)-\cos\chi\right]}H_{\mp}(\chi) &=
\sqrt{\frac{\pi}{\omega\bar{R}\sin\frac{\theta}{2}}}e^{\pm i\omega\bar{R}2\sin\frac{\theta}{2}\mp i\frac{\pi}{4}}\cos{\textstyle \frac{\theta}{2}}H_{\mp}(\chi_{\mp}
={\textstyle \frac{\pi}{2}}\pm{\textstyle \frac{\theta}{2}}) 
\nonumber
\\
	&= \mp i\frac{2\pi}{\omega\bar{R}}e^{\pm i\omega\bar{R}2\sin({\theta}/{2})}\sin({\theta}/{2})\cos^{4}({\theta}/{2})I_{R_{-}}.
\end{align}
For these values of $ \chi $ we have $c_\chi = \cos\chi_\mp =\mp\sin(\theta/2) $, $s_\chi=\sin \chi_\mp =\cos(\theta/2) $, which gives $ \psi=0 $, so Eq.~(\ref{IRm}) becomes
$ I_{R_-}= -({2v^{2}}/{\omega^{2}})    \partial_{R_{-}}F(0)$. 
Replacing all these results in Eq.~(\ref{Delta11apen}), we have so far
\begin{equation}
	\Delta^{(1,1)}=\frac{8a^{2}\omega^{2}}{3R_{*}^{5}l_{*}^{2}}\int_{0}^{\infty}d\bar{R}\bar{R}^{6}\int_{0}^{\pi}d\theta\sin\theta
	\sin{\textstyle \frac{\theta}{2}}\cos^{4}{\textstyle \frac{\theta}{2}}e^{-I_{\mathrm{tot}}}G
	\sin[\omega\bar{R}2\sin{({\theta}/{2})}]\bar{P}(c),
	\label{Deltasofar}
\end{equation}
with 
$ G=\exp\left[-\left({\omega a}/{l_{*}}\right)^{2}\sin^{2}(\theta/2)\bar{R}^{2}\right]$.

Now, for the integration with respect to $ \theta $, we shall use the  approximation (\ref{aproxlgrande}),
$ \bar{P}(c)=\sqrt{\frac{2}{\pi l_{*}\sin\theta}}\cos\frac{\theta}{2}\cos\left(l_{*}\theta-\frac{\pi}{4}\right) $. 
This is a good approximation except very close to the point $ \theta=0 $, but the integrand in Eq.~(\ref{Deltasofar}) vanishes at this point anyway.
We combine the strongly oscillating functions in (\ref{Deltasofar}) and in $ \bar P $, 
\begin{multline}
		\sin[\omega\bar{R}2\sin(\theta/2)]\cos\left(l_{*}\theta-\pi/4\right) =
		\\
		\frac{1}{2}\left\{
		\sin\left[\omega\bar{R}\left(2\sin(\theta/2)+ c_0 \theta\right)-\pi/4\right]
		+\sin\left[\omega\bar{R}\left(2\sin(\theta/2)-c_0\theta\right)+\pi/4\right]\right\} .
\end{multline}
where $ c_0 = l_{*}/\omega\bar{R} \sim 1$.
As $ \theta $ varies between $ 0 $ and $ \pi $, the first sine is highly oscillatory for $ \omega R_* \gg 1 $, and therefore is of higher order in $ 1/\omega R_* $.
Hence, we drop this term. We have to evaluate the imaginary part of the integral
\begin{equation}
	I_\theta=\int_{0}^{\pi}d\theta e^{i\omega\bar{R}\left[2\sin(\theta/2)-c_0\theta\right]+i\frac{\pi}{4}}K(\theta),
\end{equation}
with 
\begin{equation}
	K=\sin\theta\sin(\theta/2)\cos^5(\theta/2)e^{-I_{\mathrm{tot}}}G\sqrt{\frac{2}{\pi l_{*}\sin\theta}}
\end{equation}
For $  0<\theta<\pi $ the exponent  has a stationary point at $\theta = \theta_0 $ such that 
\begin{equation}
	 \cos(\theta_0/2) =  c_0 =  l_{*}/\omega\bar{R} , 
\end{equation} 
which exists only if $ \omega\bar{R}>l_{*} $ (otherwise the integral is of higher order in $ 1/\omega R_* $, so to this order we have a Heaviside function).
Thus, we   obtain 
\begin{equation}
I_\theta =	\sqrt{\frac{4\pi}{\omega\bar{R}s_{0}}}e^{i2\omega\bar{R}\left(s_{0}-c_0\arccos c_{0}\right)}K(\theta_{0}) \Theta(1-c_0).
\end{equation}
where $ s_0=\sqrt{1-c_0^2} $. Taking the imaginary part and replacing in (\ref{Deltasofar}), we obtain 
\begin{equation}
	\Delta^{(1,1)}=
	\frac{8a^{2}l_{*}^{3}}{3R_{*}^{5}\omega^{4}}\int_{l_*/\omega}^{\infty}d\bar{R}s_{0}
	e^{-I_{\mathrm{tot}}}G\sin\left[2\omega\bar{R}\left(s_{0}-c_0\arccos c_{0}\right)\right],
\end{equation}
with $ G=\exp\left[-(\omega a/l_{*})^{2}s_{0}^{2}\bar{R}^{2}\right] $ and 
$ I_{\mathrm{tot}}=\left(2\pi\bar{R}^{3}/3R_{*}^{3}\right)\left(2+3s_{0}-s_{0}^{3}\right) $.

Finally, we integrate by parts the last integral and we obtain, to lowest order in $ 1/\omega R_* $), 
\begin{equation}
	\Delta^{(1,1)}=\frac{4a^{2}l_{*}^{3}}{3R_{*}^{5}\omega^{5}}e^{-I_{\mathrm{tot}}}G
	\cos\left[2\omega\bar{R}\left(s_{0}-c_0\arccos c_{0}\right)\right],
\end{equation}
where the cosine $ c_0 $ is evaluated at the limit of integration $ \bar R=l_*/\omega $, i.e., we have $ c_0 =1 $, $ s_0=0 $, and $\arccos c_0 =0 $.
This gives Eq.~(\ref{Delta11final}).

\bibliographystyle{jhep}
\bibliography{papers}

\end{document}